\documentclass[journal]{IEEEtran}

\ifCLASSINFOpdf
\else
\fi

\usepackage[centertags]{amsmath}
\usepackage{graphics}
\usepackage{graphicx}
\usepackage{epstopdf}
\usepackage{grffile}
\usepackage[dvips]{epsfig}
\usepackage{epsf}
\usepackage{here}
\usepackage{amsthm}
\usepackage{amsfonts}
\usepackage{amsmath}
\usepackage{amssymb}
\usepackage{verbatim}
\usepackage{soul}
\usepackage{color}
\usepackage{appendix}
\usepackage{caption}
\usepackage[ruled,vlined,linesnumbered]{algorithm2e}
\theoremstyle{remark}

\newtheorem{lemma}{Lemma}

\allowdisplaybreaks
\usepackage{multirow}
\usepackage{subfigure}
\begin{document}
\title{Self-X Design of Wireless Networks: Exploiting Artificial Intelligence and Guided Learning}

\author{
}



%
\author{\IEEEauthorblockN{Erma Perenda, 
Samurdhi Karunaratne, 
Ramy Atawia, 
and Haris Gacanin
}

\IEEEauthorblockA{
Nokia Bell Labs, Copernicuslaan 50, 2018 Antwerp, Belgium
\\ 
Email: name.surname@nokia-bell-labs.com}
}

%


\maketitle

\begin{abstract}
\boldmath
In this work, we develop a framework that jointly decides on the optimal location of wireless extenders and the channel configuration of extenders and access points (APs) in a Wireless Mesh Network (WMN). Typically, the rule-based approaches in the literature result in limited exploration while reinforcement learning based approaches result in slow convergence. Therefore, Artificial Intelligence (AI) is adopted to support network autonomy and to capture insights on system and environment evolution. We propose a Self-X (self-optimizing and self-learning) framework that encapsulates both environment and intelligent agent to reach optimal operation through sensing, perception, reasoning and learning in a truly autonomous fashion. The agent derives adequate knowledge from previous actions improving the quality of future decisions. Domain experience was provided to guide the agent while exploring and exploiting the set of possible actions in the environment. Thus, it guarantees a low-cost learning and achieves a near-optimal network configuration addressing the non-deterministic polynomial-time hardness (NP-hard) problem of joint channel assignment and location optimization in WMNs. Extensive simulations are run to validate its fast convergence, high throughput and resilience to dynamic interference conditions. We deploy the framework on off-the-shelf wireless devices to enable autonomous self-optimization and self-deployment, using APs and wireless extenders.
\end{abstract}


\IEEEpeerreviewmaketitle

\section{Introduction}\label{sec:intro}

Wi-Fi self-organizing network (Wi-SON) has been proposed to proactively address different optimization challenges in dense wireless networks such as channel assignment, coverage, user control etc. \cite{Ga17} -- \cite{Chi10}. In essence, Wi-SON is monitoring network performance and calculating an optimal configuration to determine a new recommendation policy on single or clustered APs. This method, however, is deemed sub-optimal as it overlooks both internal and external network dependencies. The internal dependency refers to the relation between configurations of the AP-EXT-User set (e.g. the optimality of channel assignment depends on the location of AP, extender and end user). The external dependency appears in multi-operator deployments due to the stochastic changes of neighbor configurations  adopting the same or overlapping channels. While most efforts in SON literature \cite{Ga17,Chi10,Alim16} have been directed to define cost functions with deterministic (rule-based) optimization schemes, the above dependencies have to be explicitly addressed. In contrast, Artificial Intelligence (AI) with Machine Learning (ML) techniques should be considered to enable wireless systems with learning and sophisticated decision-making \cite{JiZh17}. To that end, we envision a truly autonomous wireless network that is capable of sensing and perceiving its neighborhood to learn network dependencies, build the necessary knowledge and enable its constituent nodes to reason out the optimal configuration. Such a design leads to the Self-X (self-configurable, self-optimizing, self-learning and self-sensing) space that allows nodes to adapt, communicate and reshape its goals based on sensed user activities. 

In this work, an AI-framework design is presented to support the network with autonomy to capture insights on system and environment evolution. We prove that our problem is NP-hard and introduce heuristics demonstrating the effectiveness of the so produced AI-driven self-optimization. The performance is validated through extensive packet-level ns-3 simulations and an experiment with commercial off-the-shelf (COTS) access points (APs). We demonstrate efficient convergence times, and verify its superiority over the state-of-the-art, before portraying its adaptability to dynamic network conditions. 
Our main contributions are as follows:

1) We initially demonstrate the tight coupling between channel and location optimization in practical multi-AP networks. The preliminary conclusions illustrate that sensing the neighborhood allows the network to pick a new setup---yet perception and learning---overlooked in the literature, remain crucial to minimize the negative impact of self-optimization on user experience, such as service interruptions.

2) We propose an AI-driven Self-X framework called Intelligent Channel Assignment and Location Optimization (ICALO) that comprises both environment and intelligent agent. The environment includes  managed APs, user devices and multi-radio wireless extenders, all modeled by a directed acyclic graph. The model considers the correlation between location and channel configurations to optimize an end-to-end user performance capturing the states of all links constituting the path from AP to a user. The intelligent agent perceives the environment by network parameters and stores them in a knowledge base (KB) that guides the learning and decision making. On the contrary, existing multi-hop optimization strategies assume the existence of non-managed neighboring information, ignore real-time performance tracking, and does not leverage the observed impact of previous actions while deriving future decisions.

3) A guided reinforcement learning (G-RL) approach is proposed with embedded domain knowledge to achieve user-aware self-optimization. The agent strikes a balance between exploration when learning has low cost, and exploitation when network performance is critical. Both perceived network states and KB are used either to select or assess new optimal configurations and retain them in the KB. The agent is aware of the learning cost that interrupts user connectivity, and thus exploits spectral correlation to transfer knowledge among matching configurations.


\subsection{Related Works}
One comprehensive survey on channel assignment in multi-radio WMNs classifies different techniques based on type of decision making, network dynamics, granularity, communication layers and optimization methods \cite{Alim16}. The decision making can be either ($i$) centralized---maintaining awareness of the network topology and state---or ($ii$) distributed, failing to maintain connectivity. A dynamic channel assignment, compared to a static one, provides a robust solution that is aware of configuration changes due to users' mobility and reconfiguration of neighboring APs. The granularity of the channel assignment is defined either at the link-level or flow-level. The former assigns the same channel to two nodes to maximize the throughput of their inter-connecting link. The latter assigns the same channel to all nodes on the flow from the source to destination. In this way, end-to-end performance is optimized without exploiting multi-radios, in which the flow can involve multiple channels while maintaining connectivity through a common channel between each two neighboring nodes. In addition, the inter-dependence such as those between links of the same flow, and between the radios of the same node, were ignored. Cross-layer channel assignment (e.g. network, data link and physical) provides a globally optimal performance, but updating routing tables and channels \cite{raniwala2005architecture} is practically infeasible with off-the-shelf devices. In addition, the neighboring interference and real-time measurements that assess the network connectivity are overlooked.

The reinforcement learning scheme in \cite{gummeson2009adaptive}---designed for sensor networks---adopted random exploration and simple reward exploitation. This can be sufficient for the considered radio and power selection problem under the foreseen slow dynamics. However, channel assignment and learning in multi-radio WMNs comprise more states and dynamics which slow down the convergence of purely random exploration, and impact the optimality of simple reward functions that do not exploit problem structure. In \cite{ding2013channel}, an Adaptive Dynamic Channel Allocation (ADCA) algorithm was proposed to pick the configuration that maximizes throughput and minimizes the delay. Each two neighboring nodes negotiate to select their common link channel that maximizes the throughput. However, the algorithm might perform sub-optimally in the case of saturated traffic and also overlooks neighboring non-managed interference (external interference).  

Finally, there are optimization techniques adopting graph coloring, integer linear programming (ILP) or meta-heuristic techniques \cite{tabuSearch, genetic}. The primary drawback of graph-coloring is its sensitivity to centralized knowledge, which usually fails to capture the granularity of inter-AP interference in non-managed scenarios. Although ILP techniques can reach globally optimal channel assignments, they fail to obtain real-time solutions in dynamic environments, and hence is not resilient. On the contrary, meta-heuristic techniques can provide near-optimal channel assignments that cope with dynamic environments, but their performance was not tested in non-managed environments. Genetic Algorithm \cite{genetic} and Tabu search \cite{tabuSearch} are considered as quasi-static searching algorithms, but they do not provide good performance in dynamic non-managed environments. CLICA \cite{Mar05} provides a channel assignment that guarantees connectivity and low inter-channel interference, but it also is not designed to handle external interference in non-managed environments. The methodology for self-deployment was presented to increase the chance of reaching optimal position of extenders at low searching and learning costs in \cite{AIGC}.


\subsection{Motivation}
The channel assignment schemes above neglect the following practical aspects: 

\textit{Neighboring network interference:} As a CSMA-based system, a target Wi-Fi station (i.e. AP, extender and user device) suffers from both exposed-node and hidden-node problems. The former refers to the contention due to neighboring nodes with high received power, operating on overlapping channels---causing busy channels and delaying transmissions. On the contrary, hidden nodes will cause packet collision at the receiver due to mutual transmission of stations outside the sensing range of each other. However, calculating the exact amount of interference and/or contention is very challenging, as the traffic profiles of non-managed neighbors are not readily available and cannot be directly predicted.

\textit{Dynamics involving extenders:} While users have the flexibility to reposition extenders, the sources of dynamics should be extended beyond user devices to include extender locations as well. Thus, a natural need arises for dynamic optimization approaches to cope with the evolution in network topology, user association and radio conditions. Such approaches should jointly solve for both channel and location of extenders to avoid positions where channel assignment is very challenging (e.g. due to excessive contention), possibly where no channel assignment is likely to offer satisfactory end-user experience. Additionally, it will mitigate the burden of moving the extender from an optimal location because of a poor channel configuration.

\textit{Learning Cost:} Both neighboring interference and network dynamics are captured through measurements performed by APs and extenders, and thus typically require channel switching and extender repositioning. Both, however, will increase the learning cost due to the service downtime due to re-association process, and the physical movement to re-position the extender. Ignoring this cost will result in poor customer experience and increased user complaints. 

\begin{figure}
\centering	
   \subfigure[Measurements of channel and location coupling in multi-hop WMN.] {
    \includegraphics[width=0.9\linewidth]{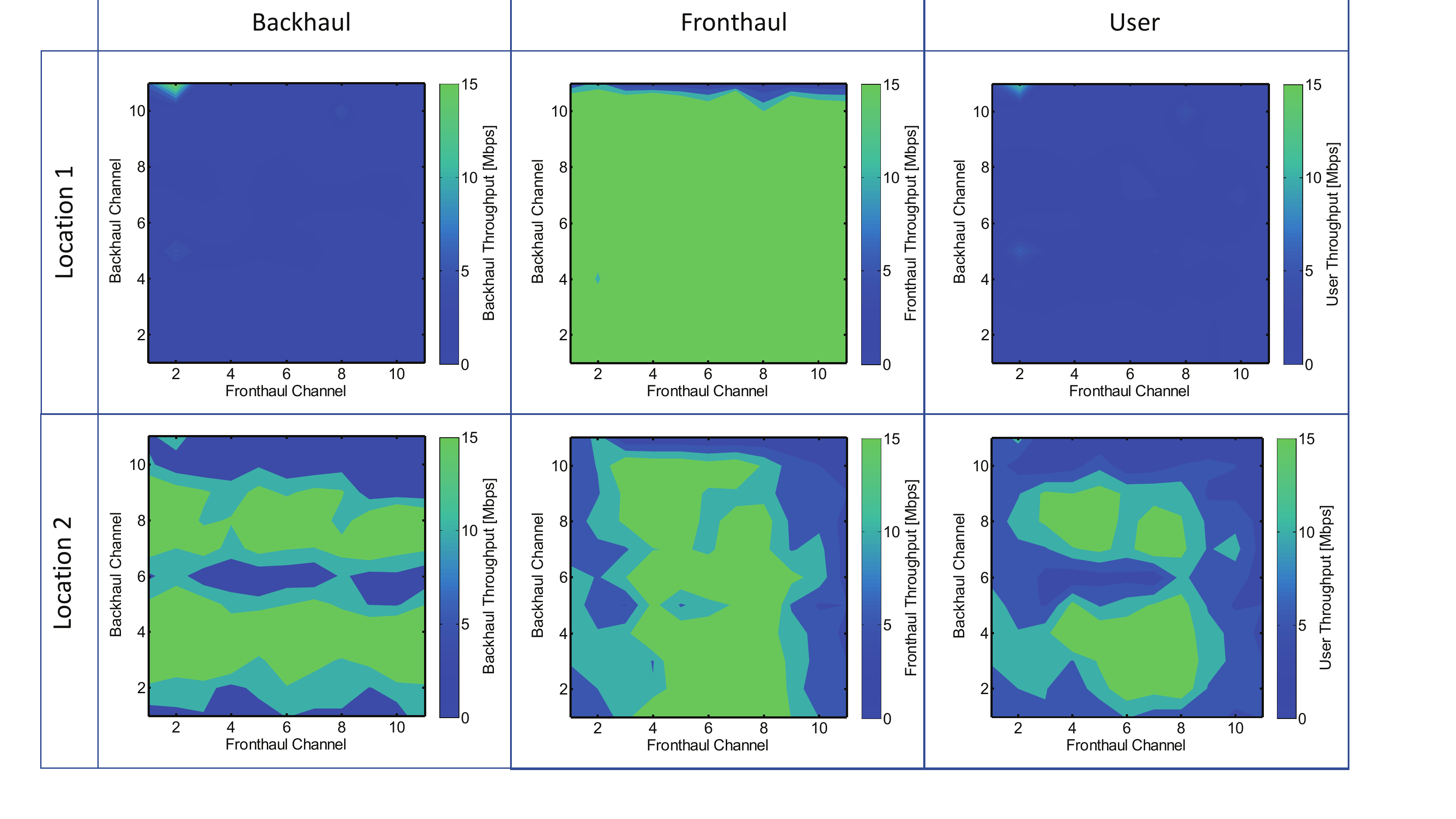}
    }
    \subfigure[A layout of tested environment.]{
    \includegraphics[width=0.9\linewidth,trim={4cm 3cm 6cm 1.5cm},clip]{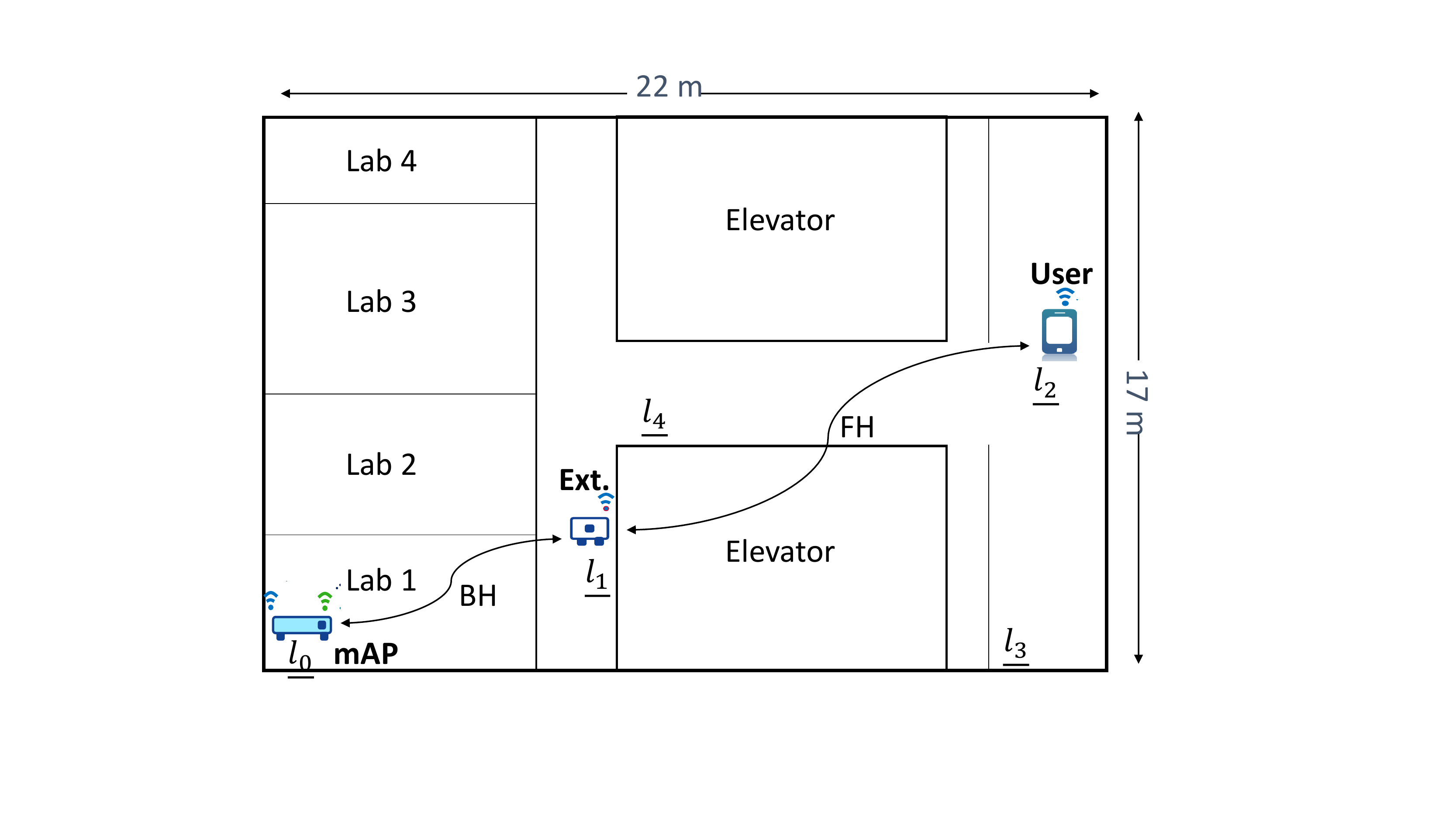}
    }
\caption{Motivation of optimization and placement problem.}
\label{fig:layout}
\end{figure}


Fig. \ref{fig:layout} illustrates a laboratory experiment of a 2-hop WMN with a single user connected to an AP through one extender. We denote the link between AP and extender by backhaul (BH) and the link between a user device and extender by fronthaul (FH). The throughput  at the BH, FH and user device at two different extender locations $l\in\{l_1, l_4\}$, is measured and denoted by $ R_b^{(l)} $, $ R_f^{(l)} $ and $ R_u^{(l)}=min(R_b^{(l)},R_f^{(l)})$, respectively. At the first location (location $l_4$ in Fig. \ref{fig:layout}), the BH throughput (and hence the user throughput) is maximized only by selecting the channel combination 2 and 11 for fronthaul and backhaul, respectively. Such a location is said to be sub-optimal for the backhaul as it suffers from high interference and/or low coverage. On the contrary, at location $l_1$, $ R_b $ is optimized over a wide range of channel combinations while  $ R_f $ is maximized over a more tighter range of optimal channel combinations and do not include channels 2 and 11 that were deemed optimal at the first location. 
The network should be aware that the deployment is sub-optimal \cite{AIGC} in the first location and performance improvement through channel assignment is not attainable, while the likelihood of reaching an optimal channel combination is very high in the second location. An unprecedented challenge is reducing the cost of learning which was very high at the first location due to the poor BH. As such, a delay of up to tens of seconds was experienced to collect BH measurements, followed by an extra delay of a few seconds to re-associate the user with the extender and the extender with the AP. Addressing these challenges is goal of this work.

		

\section{Network Model}\label{sec:related}
We consider a semi-stationary multi-radio multi-channel (MR-MC) WMN with a single (static) master AP (mAP) and \textit{M} wireless extenders (EXTs), whose locations change over time. The mAP is a gateway which provides access to the Internet, while EXTs act as relays to extend the Wi-Fi service region of the mAP. The extenders and mAP are equipped with a number of radio interfaces, where each radio is operating on one channel. The network is serving \textit{U} user devices that are connected to the mAP either directly or through extenders.

We model the network as a directed acyclic graph $G=(V,E)$, where \textit{V} is the set of nodes and \textit{E} is the set of bidirectional links (edges). $v_i \in V$ represents either mAP, EXT or user device, where $v_0$ refers to the mAP, $v_{1},\ldots,v_M$ represent the extenders, and $v_{M+1},\ldots, v_{M+U}$ are user devices. We assume \textit{N} available channels and \textit{L} possible locations for deploying extenders. We denote the set of radio interfaces for each node $v_i$ by  $D_{i}$, and the set of channels associated to radio-interfaces by $C_{i}$. Each link $e_{ij}\in E$ comprises of two nodes $v_i$ and $v_j$, where $v_j$ is connected to $v_i$ and the latter provides the next hop communication to the mAP. Both nodes are in the transmission range of each other and they have at least one common channel assigned to their interfaces (i.e. $C_{i}\cap C_{j}\neq \varnothing$). We define $h_{ij}$ as the channel associated with $e_{ij}$ and thus, the link can be represented with a triple $e_{ij}=\{v_i, v_j, h_{ij}\}$. 



\subsection{System Variables}
The system variables of our implementation model are described in the following.

\subsubsection{Links and Paths} 
We define the $k$-th user path $p_{k}=\{e_{ij} \mid i,j \in [0,M+U]; i \neq j\}$ as a set of distinct links $e_{ij} \in E$ connecting mAP $v_0$ and the $k$-th user node $v_k$ ($k \in [M+1, M+U]$ denotes user index). We constrain two successive links $e_{ij}$ and $e_{nm}$ in path $p_{k}$ by setting $j=n$. The set of nodes forming the links of path $p_{k}$ must contain only one node each with index 0 and index $k \geq M+1$.

\subsubsection{Location-specific RSSI}
The Received Signal Strength Indicator (RSSI) at receiver node $v_j$ at location $l_{j}$ from sink node $v_i$, ${RSSI}_{ij}^{(l_{j})}$ presents a measured received signal strength in dBm of beacon frames received on the channel (i.e. defined as dot11BeaconRssi \cite{defUtilization}). RSSI is usually measured during the reception of the physical (PHY) preamble and its value is forwarded to the Medium Access Control (MAC) layer in the RXVECTOR \cite{defUtilization}. Beacons' RSSI may be averaged over time using a vendor specific smoothing function. In case that the beacon frame is received by means of multiple receive chains, the RSSI is averaged in linear domain over all chains. The RSSI value range is -100 dBm to 40 dBm \cite{defUtilization}.

\subsubsection{Channel Utilization} 
The utilization $u_{{ij}} ^{(l_{j})}$ for channel $h_{ij}$ at location $l_{j} \in L$ is defined as the fraction of time that the channel was busy, linearly scaled with 255, as indicated by either the physical or virtual carrier sense mechanisms \cite{defUtilization}. The utilization is calculated at the AP level as follows 
$u_{{ij}}^{(l_{j})}=\lfloor\frac{CBtime(t)}{BI\times \Delta BI \times 1024}\times 255\rfloor$, 
where $CBtime(t)$ and $\lfloor\cdot\rfloor$, respectively, denote the channel busy time at time-instant $t$ measured during the Clear Channel Assessment procedure and the math floor function. $CBtime$ is defined as the total time duration (in microseconds) where the carrier sense mechanism observed the channel to be busy between predefined beacon intervals. It is an implementation specific parameter usually measured as a summation of the duration between start of the $PHYCCA.indication (BUSY) primitive$ and start of the $PHYCCA.indication (IDLE) primitive$.  \textit{BI} represents the number of consecutive beacon intervals during which the channel busy time is measured (i.e. Dot11ChannelUtilizationBeaconIntervals in IEEE 802.11 \cite{defUtilization}) and $\Delta BI$  is the duration between two successive beacons.

\subsubsection{Link Throughput} 
We define throughput of the link $e_{ij}=\{v_i,v_j,h_{ij}\}$ at receiver node $v_j$ placed at location $l_{j} \in L$, as $R_{{ij}}$.  The maximal link throughput is obtained as follows \cite{defUtilization}:
\begin{eqnarray}\label{eq:linkthroughuput}
R^{max}_{{ij}}=\min{(\log_{2}{(1+10^{\frac{{RSSI}_{ij}^{(l_{j})}+P_{adjust}}{10}})}, maxBPS)} \times  \\
\frac{maxNSS}{PPDU}\times N_{OFDM},\nonumber
\end{eqnarray}
where $P_{adjust}$ is the implementation specific power adjustment parameter in dBm taking into account potential transmit power differences between Beacon/Probe response frames to data frames; $maxBPS$ denotes a maximum number of bits per second which is equal to: 40/6 if 256-QAM 5/6 modulation is allowed in the link, 6 if 256-QAM 3/4 modulation is allowed in the link or to 5 otherwise; $maxNSS$ is the maximum number of spatial streams; $N_{OFDM}$ denotes the number of OFDM sub-carriers and $PPDU$ is the duration of one physical protocol data unit payload symbol in seconds \cite{defUtilization}. The link throughput value is calculated for transmit and receive modes whose values are stored as an L\_DATARATE parameter within TXVECTOR and RXVECTOR primitives \cite{defUtilization}. For example, these values can be obtained through Broadband Forum Technical Report (TR)-181 specification as \textit{InternetGatewayDevice.LANDevice.\{i\}. WLANConfiguration. \{i\}.AssociatedDevice.\{i\}.X\_BL\_TxRate} and \textit{InternetGatewayDevice.LANDevice.\{i\}. WLANConfiguration.\{i\}.AssociatedDevice.\{i\}.X\_BL\_RxRate} \cite{tr181}. The maximum link throughput is multiplied by the percentage of time the medium is idle to obtain the link throughput given by
\begin{equation}\label{eq:linkthroughuputmodified}
R_{{ij}}^\prime=R^{max}_{{ij}}\times (100-u_{{ij}}^{(l_{j})}).
\end{equation}

\subsubsection{End-to-end Throughput} 
Computationally, the end-to-end throughput of the $k$-th user device is defined as $R_k=\min\{R_{{ij}} \mid e_{ij} \in p_{k}\}$. On the other hand, $R_k$ can be practically estimated as a user goodput based on transmitted and received bytes by the user within a measurement period $\Delta t$ as 
$R_k=\frac{(TXBytes + RXBytes)\times 8}{\Delta t}$, 
where $TXBytes$ and $RXBytes$, respectively, denote the total number of bytes transmitted and the total number of bytes received. These values are available through specific vendor extensions (e.g. statistics counters \textit{InternetGatewayDevice.LANDevice.\{i\}.WLANConfiguration.\{i\}.AssociatedDevice. \{i\}.Stats.BytesSent} and \textit{InternetGatewayDevice.LANDevice. \{i\}.WLANConfiguration.\{i\}.AssociatedDevice.\{i\}. Stats.Bytes Received}, respectively). 
Although the second way to obtain  end-to-end user throughput is more accurate than the first, it has one drawback since it requires that the user-devices are always active with the transmitting and receiving data requests. 

\subsection{Problem Formulation} 
We define the objective function of our approach as the total end-to-end throughput of all user devices, written as 
$\max_{C,L} \sum_{k=M+1}^{M+U} R_k$, 
where the search is done across a set of channels $C=\{C_{i} | i=[0,M]\} $ and a set of locations $L=\{\l_{i}| i=[1,M]\}$ that lead to optimal network configuration for each path $p_k, k=[M+1,M+U]$. Optimization of the objective function is done under the following constraints:

(a) Finite set of available channels -- the set of channels that can be assigned to any node is $N$.

(b) Channel-radio relationship -- to each radio can be assigned only one channel. That is $\forall v_{i}\in V, card(C_{i}) = card(\mathcal{D}_{i})$, where $card(\cdot)$ denotes the cardinality of a set.

(c) Radio constraints -- the number of channels assigned to one node cannot exceed the number of radios on the node. That is $\forall v_{i}\in V, \bar{\mathcal{C}}_{i}\leq  \bar{\mathcal{D}}_{i}$, where $\bar{\cdot}$ denotes the number of distinct elements in a set---which means that the same channel can be assigned to different radios of $v_i$.

(d) Connectivity -- two adjacent nodes $v_i$ and $v_j$ must have at least one channel in common $\mathcal{C}_{i}\cap\mathcal{C}_{j}\neq\varnothing$.


The computational hardness property of the above defined objective function is provided by the following Lemma.

\begin{lemma}
Joint channel assignment and location optimization in WMNs possess the non-deterministic polynomial-time hardness (NP-hard) property.
\end{lemma}

The proof of Lemma 1 is given in Appendix A. We note here that, unlike the colouring problem formulation given in the proof, our problem considers a fully dynamic neighbouring environment and search for an optimal configuration set of channel and location. Hence, below we present a heuristic algorithm with guided learning to achieve a near-optimal configuration.

\section{Self-optimization Framework Design}

\begin{figure}
\centering
	\includegraphics[width=0.9\linewidth,trim={3cm 11cm 18cm 2cm},clip]{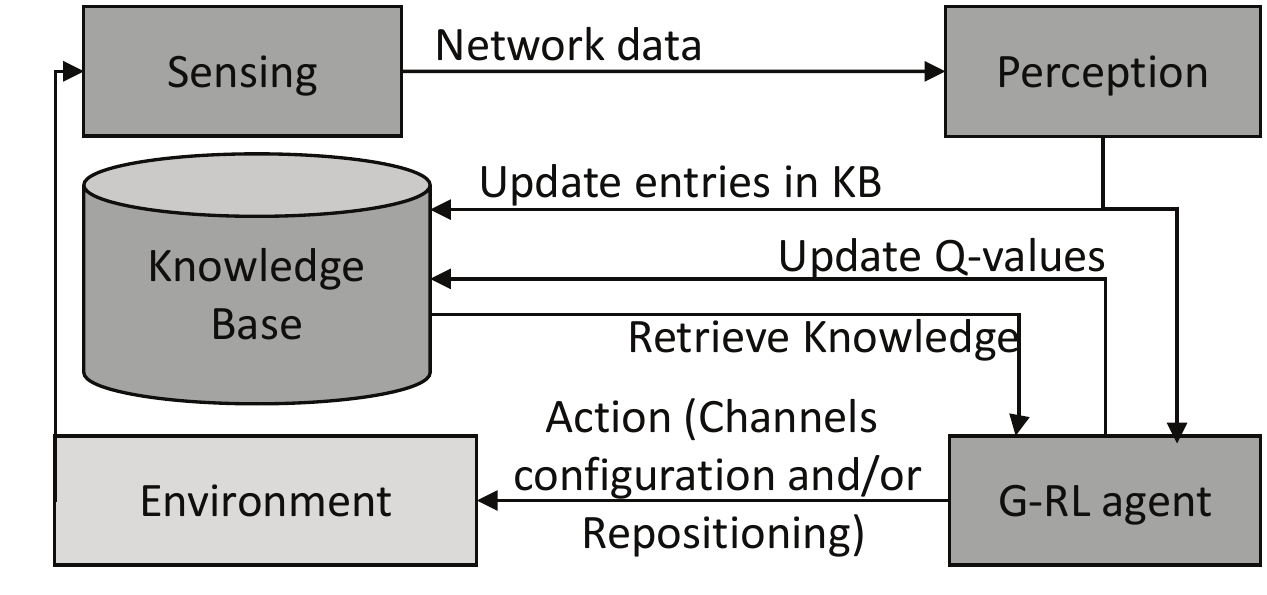}
\caption{AI-driven self-optimization framework.}
\label{fig:framework}
\end{figure}

A key aspect of self-optimization is the autonomy, in which the network can configure both the mAP and extenders without manual troubleshooting or instructions by the operator's help desk. The AI is thus adopted to enable learning, perception and reasoning which supports the network with the knowledge necessary for autonomous decision making \cite{AIBook}. The network is typically modeled as two main elements: Environment and Intelligent agent. The former consists of a managed Wi-Fi system (mAP and its extenders) and non-managed neighboring APs. The agent interacts with the environment by sensing the current state and then provides actions by reinforcement learning (RL) \cite{RLBook}. The agent then evaluates the actions based on a certain reward, which is a function of the resultant network state. It stores the perceived states and rewards of each action in a KB that can be utilized to improve the quality of future decisions. The overall architecture of the proposed AI framework is summarized in Fig. \ref{fig:framework}, and comprises the environment, the KB and their interaction with the agent: sensing, perception and reinforcement-learning.


\subsection{Sensing}
In the sensing stage, the values of physical parameters that can be used to describe the Wi-Fi system may be collected from the mAP and EXTs (i.e. from each $v_{i} \in V, i \in [0,M]$) through the TR-98/181 protocol for remote management \cite{tr181} or through another interface defined by the device software development kit. The collected information contains radio-interface level statistics (e.g., the indices of used channels, Clear Channel Assessment stats counters such as channel busy time etc.) and user-device level statistics (e.g. RSSI, counter values for total number of retries, failed packets, sent packets, sent and received bytes etc.). This information will allow the agent to perceive the environment, detect its current state and assess the performed actions. The sensing stage collects the data from each node with a certain period $\tau$ in milliseconds.

\subsection{Perception} 
 The perception phase translates the sensed information from each node $v_{i}$ into performance indicators (i.e. system variables) that identify the network state. The performance indicators are calculated for each radio $d \in D_{i} \text{ of node } v_i \in V, i \in [0,M]$ based on two successive sensing samples. These indicators include:

\textit{Channel utilization} $u_{{ij}}^{(l_{j})}$ (in \%) is very rarely provided by the chipsets and it is usually calculated manually within the time interval $\tau$ based on the Clear Channel Assessment stats counters (provided by the chipsets) of channel busy time in milliseconds as 
$u_{d}=\frac{CBtime(t+\tau)-CBtime(t)}{\tau} \times 100$.

\textit{Activity factor} (in \%) is calculated based on Clear Channel Assessment stats counters channel transmit time (\textit{CHTXtime}($\cdot$)) and channel receive time (\textit{CHRXtime}($\cdot$)). Each of the above mentioned Clear Channel Assessment statistics parameters are vendor implementation specific---however, they are calculated based on different Clear Channel Assessment  and PHY states indications (i.e. BUSY, IDLE, TX, RX). Accordingly, the activity factor is given by
$\rho_{d}=[\frac{CHRXtime(t+\tau)+CHTXtime(t+\tau)}{\tau}-\frac{CHRXtime(t)+CHTXtime(t)}{\tau}]\times 100$, 
where \textit{CHRXtime}($\cdot$) and \textit{CHTXtime}($\cdot$), respectively, denote the total time in milliseconds that the radio has spent on receiving data and the total time in milliseconds it has spent on transmitting data. The values of these counters might be obtained during the PHY receive and PHY transmit procedure. \textit{CHRXtime}($\cdot$) is calculated as summation of the periods between PHY-RXSTART indication and PHY-RXEND indication, while \textit{CHTXtime}($\cdot$) is calculated as a summation of the periods between PHY-TXSTART  indication and PHY-TXEND indication \cite{defUtilization}.

\textit{Retries rate per user device} (in \%) is calculated based on user-level statistics data as 
$\Delta_{retr,k}=\frac{N_{retr,k}(t+\tau)- N_{retr,k}(t)}{N_{pack,k}(t+\tau)N_{pack,k}(t)} \times 100$,  
where $N_{retr,k}(t)$ is the total number of retries for the $k$-th user device  at time instant $t$ (e.g. vendor-specific implementation \textit{InternetGatewayDevice. LANDevice.\{i\}. WLANConfiguration.\{i\}. AssociatedDevice.\{i\}. X\_BL\_TxRetries}) and $N_{pack,k}(t)$ is the total number of packets transmitted out of the interface for the $k$-th user device at time instant $t$ given by \textit{InternetGatewayDevice. LANDevice.\{i\}. WLANConfiguration.\{i\}. AssociatedDevice.\{i\}. Stats. PacketsSent} \cite{tr181}.

\textit{Error rate per user device} (in \%) is calculated as 
$\Delta_{err,k}=\frac{N_{err,k}(t+\tau)- N_{err,k}(t)}{N_{pack,k}(t+\tau)- N_{pack,k}(t)} \times 100$, 
where $N_{err,k}(t)$ is the total number of inbound failed packets for the $k$-th user device at time instant $t$ (e.g. vendor-specific implementation \textit{InternetGatewayDevice. LANDevice.\{i\}. WLANConfiguration.\{i\}. AssociatedDevice.\{i\}. Stats. X\_BL\_TxFailed}).

We note here that retries and error rate per user device give an insight to the severity of interference level. High level of interference consequently results in high values of error and retries rates for the user-devices impacted by the interference. The channel utilization metric gives an insight to the contention level, since the activity factor provides information on how much the radio traffic load contributes to that contention level. The value of the ratio of the activity factor to the channel utilization of the channel assigned to the radio  $d$ is used as a perception control variable along with the channel utilization value to trigger channel optimization. When this ratio has a very low value and the channel utilization value is higher than a certain threshold, the perception stage will detect the current network state as sub-optimal, resulting in an evaluation of the current state in the network. In order to avoid false alarms, the perception stage is responsible to correct values of the activity factor for radios that have connections among themselves. For an example, assume that the link $e_{ij}$ is formed of two nodes $v_i$ and $v_j$, where node $v_i$ is a parent node. If the parent node has some other user-devices connected to the same radio $d_{v_i}$ as node $v_j$, then a high activity factor of the radio $d_{v_i}$ of the parent node will be observed as a high channel utilization of the radio $d_{v_j}$ at node $v_j$. In case that the activity factor of radio $d_{v_j}$  has a very low value, but its channel utilization value is very high, it will consequently trigger the channel optimization, although channel switching cannot change the current state as its parent node mostly contributes to its channel utilization. This is a false alarm and it is necessary to modify the activity factor of the radio $d_{v_j}$ at node $v_j$ to the activity factor of radio $d_{v_i}$ of its parent node.

By means of real-time network monitoring of the aforementioned metrics, the perception component is capable of detecting when the current configuration becomes sub-optimal, and sending a signal to the G-RL agent (defined below) to evaluate the current state of the network.

\subsection{Reinforcement Learning}
The guided RL (G-RL) agent utilizes Q-learning to select the optimal action at each state based on stored reward values (referred to as Q-values). In essence, the G-RL agent considers that each node $\{v_{i}|\forall i \in [0, M]\}$ has its own states and corresponding actions in that state, while the rewards are derived on the system level.
The states, actions and rewards for each node $v_{i} \in V, i \in [0, M]$ are defined as follows:

\emph{States (S):} Beside channels optimization, G-RL agent aims to place each node $v_i$ at an optimal location. Thus, the state $ s \in S $ of each node will refer to its location $ l_{i} $. Each node $v_i$ has $L$ possible locations for deployment and hence $L$ possible states.

\emph{Actions (A):} G-RL agent takes two types of actions: channel configuration $ A^{(c)} $ and EXT repositioning $ A^{(l)}$ with action set $A(s)= A^{(c)}\cup A^{(l)}$. Since each node $\{v_{i}|\forall i \in [0, M]\}$ is equipped with $D_{i}$ radios, we define channel configuration actions for that node as the set of all possible combinations of the radios, where $ |A^{(c)}| = N^{|D_{i}|}$. On the other hand, each repositioning action $ a \in A^{(l)} $ changes the location of node $v_i$ and results in a state transition.

\emph{Reward (R):} Instantaneous reward at time instant $t$ in the state $s$ for a selected action $a$ at node $v_i$ is given by 
$r_t(s,a, v_{i})=\sum_{k=M+1}^{M+U}{R_k}.$ 
We define the reward at the network level because applying an action $a$ at node $v_i$ impacts performance of whole network.
In Q-learning, the cumulative reward $Q_t(s,a, v_{i})$ is calculated using the previous Q-value and the instantaneous reward \cite {RLBook} as given by
\begin{eqnarray}\label{eq:qval}
\begin{cases}
Q_t(s,a, v_{i}):=Q_t(s,a,v_i)+\eta \Delta(s,a) \\
\Delta(s,a) = r_{t}(s,a, v_{i})+ [\gamma \max_{\forall a \epsilon A} Q_{t+1} - Q_t(s,a)]
\end{cases}
\end{eqnarray}
where $Q_t(s,a)$ is the cumulative reward at state $s$ when action $a$ is applied at time $t$. Parameters $\eta$ and $\gamma$, respectively, are the learning factor and discount rate with values between 0 and 1. $\eta$ controls the convergence speed of the learning and its value is gradually decreased in time to achieve convergence. The discount rate, $\gamma$, is used to weight the near-term rewards. Specifically, as $\gamma$ approaches 1, the weight of future rewards is increased.


\emph{Policy ($ \pi $):} The selection of action $ a $ during a certain state $ s $ is governed by a policy  $\pi(a|s)$. A policy that maximizes the cumulative reward $ Q_t(\cdot) $ is denoted as $ \pi^*$. During the early stages of learning, when the KB is empty, the G-RL agent has to explore in order to discover the unknown environment. Subsequently, the KB is populated and the agent can retrieve and start exploiting the gained experience to pick an action that has the highest reward. Finding the optimal trade-off between exploration and exploitation is very challenging while deriving the policy, as it impacts both the learning cost and convergence rate \cite{RLBook}.

\subsection{Knowledge Base}
The knowledge base stores the three types of tables for each node $v_{i} \text{, } i \in [0,M]$ as shown in Fig. \ref{fig:kb}.

\emph{Perception table:} stores all the information related to the connectivity in the network and the parameters calculated in the perception phase. For each radio-interface $d \in \mathcal{D}_{i}$, this table stores all the next hop nodes, the used channel $h_{d}$ of each radio, and the changes in utilization and activity factor denoted by $u_{d}$ and $\rho_d$, respectively. With regard to connected users, the changes in retries and error rates, denoted by $\Delta _{retr.}$ and $\Delta _{err.}$, respectively, are also stored.

\emph{Q-table:} this table saves the Q-values for each possible action $ a $ in state $ s $ calculated by Eq. \ref{eq:qval} \cite{AIBook}. 

\emph{Channel-Location table:} the channel utilization of all available channels $N$ and at all candidate locations $L$ is kept at each time slot. Entries are set only for channels that were sensed at a certain location. Otherwise, the entries remain empty. 
	
With such a design of KB, the G-RL agent is aware of network topology and current status in the wireless system.

\section{Guided RL Agent Design}
The RL agent is considered as both a learner and a decision maker. Thus, the agent has to balance between exploring the environment to gain more information, and exploiting the KB by picking decisions with a high likelihood to reach the optimal state. While the user experience during such learning and decision making processes remains a priority, the RL agent has to be guided by domain experience to minimize the learning cost. To that end, problem-specific knowledge is used, instead of random exploitation and exploration, to provide a user-aware decision at the right time. 
In essence, the agent explores the environment when 1) the observed change in the reward values is insignificant, or 2) the learning cost is low due to the absence of user traffic. On the contrary, exploitation is applied when 1) large (positive) variations in the reward values are detected, or 2) interference or contention problems are perceived. During both stages, the agent is aware of the following domain knowledge: \\
\textit{Spectral Correlation}: Overlapping channels in a Wi-Fi system\footnote{Here, we consider 2.4 GHz band, but overlapping in 5 GHz band is observed by usage of dynamic channel bandwidth (20, 40, 80 and 160 MHz).} will typically have similar utilization factors since a given channel can be sensed busy due to transmission on the same or an overlapping channel. Thus, the exploration stage should pick non-overlapping channels, while overlapping channels are visited through exploitation.\\
\textit{Spatial Correlation}: A Wi-Fi system that is typically suffering from a coverage problem can not be optimized by re-configuring the channels, and thus prompts a change in the location of nodes (i.e. re-positioning EXTs). As such, identifying the coverage problem from contention and interference will help the agent to exclude channel configuration from the set of possible actions, and thus accelerate the learning process.

The main stages of ICALO are summarized in Algorithm \ref{alg:alg1} and detailed as follows:

\subsection{Selecting the Type of Action}
\begin{figure}
	\centering
	\includegraphics[width=0.9\linewidth,trim={0cm 2cm 2cm 4cm},clip]{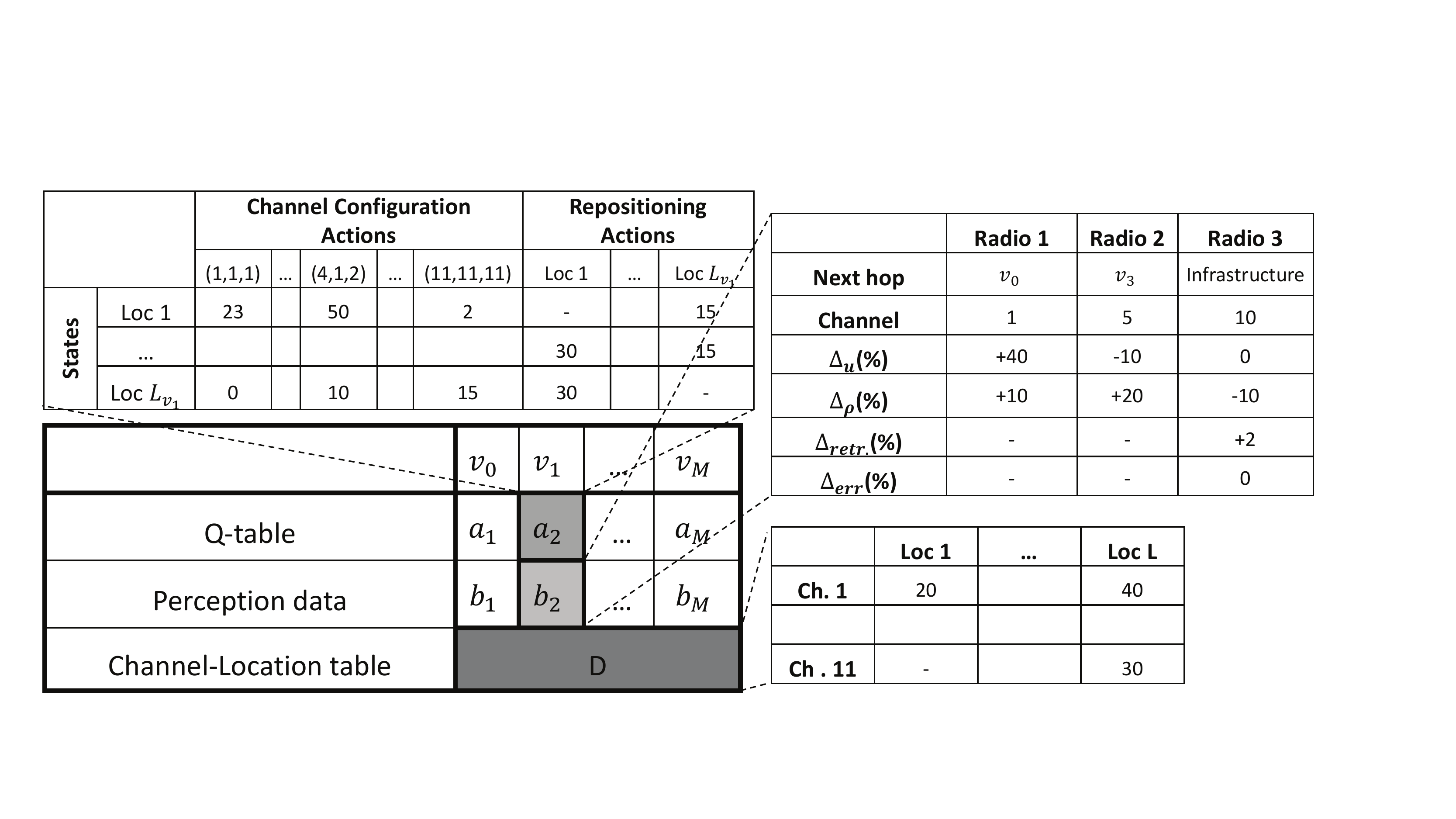}
	\caption{Knowledge Base Design.}
	\label{fig:kb}
\end{figure}

Using the perception data, the agent monitors the system performance by checking the changes in contention, interference and coverage levels at current extender location. In particular, the $ RSSI $ value on EXT's BH (connecting EXT to mAP) is assessed versus a minimal threshold $ RSSI' $ that achieves the target signal quality at the extender if the channel is optimized (Lines 4-6). In the case of poor coverage, channel exploration at such a location is unnecessary and thus a repositioning action must be selected. The new location is calculated as the midway between the current position of the extender and the next hop towards the mAP. If this location was visited before, then a random distance is added to the calculated midway location to provide exploration. The new location is stored in the channel-location table in Fig. \ref{fig:kb} with the corresponding channel utilization of the last channel configuration.

In case of high signal level, i.e. no coverage problem, the agent should explore and exploit using the channel  configuration actions until no improvement is observed, and then a new location is selected (Lines 4-6).

\subsection{Zero-Cost Knowledge-Driven Exploration}

The second policy performs greedy exploration, yet with zero learning cost, since it is followed when 1) no users are associated or 2) the connected users are not requesting any traffic (Line 8). In particular, the agent will pick a channel configuration action, compute its reward value and store the cumulative reward in the Q-table (Lines 10-12) to maximize the gained knowledge. As such, for every possible action that is not applied before (i.e. with zero reward value in Q-table), the total Euclidean distance, to all previously visited actions, is calculated by $\beta_a$ as a sum of Euclidean distances between action $ a $ and the previously applied actions $ I $ stored in the Q-table. $ c_{a,d} $ is the channel configuration of radio $ d $ when applying action $ a $.
The optimal action, from the exploration perspective, is the one with maximum total Euclidean distance.

In the case that all actions are visited (i.e. no zero entries in the Q-table), a random action is picked from the Q-table using a uniform distribution (Lines 14-16). This exploration process is repeated for every node $\{v_i | \forall i \in [0,M]\}$ until a connection or traffic request is received from a user device. By doing so, the G-RL agent accelerates the learning process of its environment without the degradation of user experience. After the agent applies this exploration action, the corresponding Q-value is updated in KB, and the channel configuration is switched back to the former value (Line 20).


\subsection{Modified Basic Soft Max: Exploiting Spectral Correlation}
In the case of perceiving interference or contention problems, the third policy is triggered (Lines 24-38). In essence, the third policy is defined based on Basic Softmax (BSmax) combined with Value-Difference Based Exploration (VDBE – Softmax) \cite {vdbeSoftmax}, spectral correlation and KB.

\begin{equation}
\pi (s)=\left\{
\begin{array}{ll}
\text {Modified BSmax policy} & \xi<\varepsilon (s)\\
arg \max_{a\in A(s)}Q(s,a) & otherwise,
\end{array}
\right.
\end{equation}
where $\xi$ is a uniform random number over the interval [0, 1], and $\varepsilon (s)$ is a state-dependent exploration probability. In essence, {a high value of $\varepsilon (s)$ enables the agent to perform guided exploration, while a low value triggers exploitation by picking the action with maximum cumulative reward (i.e. Q-value)}.

\subsubsection{Exploration Probability $\varepsilon(s)$}
The state-dependent exploration probability $\varepsilon (s)$ is calculated using the difference in Boltzmann distribution between the last two cumulative rewards:
\begin{equation}
\label{eq:vareps}
\left\{
\begin{array}{ll}
\varepsilon_{t+1}(s)=\psi(s) f(s,a,\sigma)+[1-\psi(s)] \varepsilon_t (s), \\
f(s,a,\sigma)
=\frac{1-e^{\frac{-|\eta\Delta (s,a)|}{\sigma}}}{1+e^{\frac{-|\eta\Delta(s,a)|}{\sigma}}} 
\end{array}
\right.
\end{equation}
where $\sigma$ and $\psi \in [0,1]$, respectively, denote a positive constant called inverse sensitivity and the influence of the selected action on the state-dependent exploration probability. A reasonable setting for $\psi(s)$ is the inverse of the number of actions in the current state, $\psi(s)=\frac{1}{|A(s)|}$, since all actions should contribute equally to $\varepsilon(s)$. The parameter $\sigma$ influences $\varepsilon (s)$ in a way that low values cause full exploration at small value changes while high values of $\sigma$ cause a high level of exploration only at large value changes.

\subsubsection{Action Selection}
In the case of $\xi<\varepsilon (s)$, the G-RL agent is in exploration phase. The exploration phase takes five steps to pick a new action. First, the G-RL agent calculates for each $a \in A(s)$ the action selection probability $ \rho (s_t=s,a_t=a,v_i)= min \{\rho_o (s,a,v_i), \rho_u (s,a,v_i)\}$
by using the BSmax probability $\rho_o (s,a,v_i)$ and the environment probability $ \rho_u (s,a,v_i)$. The latter probability takes into account channel diversity, hidden node impact and contention impact caused by channel utilization and overlapping channels. $ \rho_o (s,a,v_i)$ is determined using a Boltzmann distribution $P_r\{a_t=a|s_t=s\}= e^{\frac{Q(s,a)}{T}}/\sum_{b \in A(s)}{e^{\frac{Q(s,b,v_i)}{T}}}$, 
where $T$ is a positive parameter called temperature starting with a large value and decreases with time. High temperatures cause all actions to be nearly equiprobable (more exploration), whereas low temperatures cause greedy action selections (more exploitation), while 
$\rho_u (s,a,v_i)=\frac{CD}{UI+HI+CI}$. 
Here, $CD$ denotes the impact of channel diversity given as 
$CD=1+\sum_{d \in D_i}{\sum_{d^{\prime} \in D_i, d \neq d^{\prime}}{|h_d-h_{d^{\prime}}|}}$ 
so that the action with the same channel tuned on all radio interfaces has the lowest probability. $UI$ denotes the impact of channel utilization on the action given as 
$UI=\sum_{d \in D_i}{u_{h_d}^{l_i}}$. 
$HI$ denotes the impact of hidden nodes and is defined as the difference between channel utilization observed on both sides of links that contains node $v_i$, multiplied by a factor 100, as 
$HI=\sum_{j, e_{ij} \in E}{|u_{h_{ij}}^{l_i}-u_{h_{ij}}^{l_j}|}\times 100$. 
Finally, $CI$ denotes the impact of contention from overlapping channels given as 
$CI=\frac{\sum_{d \in D_i}{\sum_{h \in \mathcal{N},|h-h_d|\leqslant 5, h \neq h_d}{(5-|h-h_d|) u_{h}^{(l_i)}}}}{50}$. 
The second step is finding the maximal probability $\rho_{max}=\max(\rho)$ and on basis of it calculating the minimal allowed probability as 
$\rho_{min}=0.9\times\rho_{max}.$

In the third step, the G-RL agent finds all actions $A(s)^{\prime}$ for which $\rho(s,a,v_i)>\rho_{min}$. Afterwards, the G-RL agent calculates $\kappa_{a\prime}$ for each $a\prime \in \mathcal{A}_s^\prime$ as 
$\kappa_{a^{\prime}}=\bigg(1+\sum_{d \in D_i}{\sum_{d^{\prime} \in D_i, d \neq d^{\prime}}{|h_{d,new}-h_{d^{\prime},new}|}}\bigg)\times\sqrt{\sum_{d \in D_i}{(h_{d,current}-h_{d,new})^2}}$. 
The first factor denotes the channel diversity of the new action, since the second denotes Euclidean distance from the current applied action at node $v_i$. If the applied action does not satisfy perception thresholds, then it is highly likely that the actions with low Euclidean distance behave in the same way. However, the G-RL agent gives higher probability to actions that have higher Euclidean distance from the current applied action. In the last step, the G-RL agent picks the action $a^*$ that has highest $\kappa$ value.

\subsection{Decision Making - Control Stage}
After a new action is found, ICALO checks whether it knows anything about this action, i.e. whether the Q-value for this action is different from zero (Lines 40-45). In the case that Q-value is equal to zero, ICALO will apply the new action. Otherwise, it checks whether the Q-value of the new action is 15\% higher than the Q-value of currently applied action. This is because it is not worth applying a new action if it brings only a small improvement. By controlling the execution of actions in such a way, ICALO alleviates the issue of the network oscillating between the same states.

\renewcommand{\baselinestretch}{1} 

\begin{algorithm}
\begin{small}
\caption{Guided VDBE – Softmax Q-Learning}
\label{alg:alg1}
\SetAlgoLined
\SetKwInOut{Input}{Input}
\SetKwInOut{Output}{Output}
\SetKwInOut{Initialization}{Initialization}
\Input {Knowledge Base (Q-table, Perception data and Channel-Location data)\;}
\Output  {Action $a^*$\;}
\textbf{Define:} Max. channel utilization: $u_{thr}$, max. re-transmission rate: $\Delta_{retr._{thr}}$, max. error rate: $\Delta_{err._{thr}}$, min. signal level: $ RSSI' $, and target Q-value: $ q' $\;
\For {$v_i \in V$}
{
/* Policy 1: select type of action */ \\
\If {$RSSI \leq RSSI'$ OR $max(Q)<q'$}
{
$ a^* =  OptimizeLocation$\;
}

/* Policy 2: zero-cost guided exploration */ \\
\While{$U==0$} 
{
	$ c_{old} =  c_ {inf}$\;
	\If {$ min (Q) == 0 $}
	{
		Calculate $\beta_a = \sum_{i=1}^{I} \sqrt{\sum_{d=1}^{D} ({c_{a,d}-c_{i,d}})^2}$ \;
		$ a^* = argmax \{\beta_a\} \ \forall a$ \;	
	}
	/* All actions are visited before */ \\
	\Else
	{
		$ a^* =  $ Random uniform selection \;
	}
	Apply $ a^* $ to $ c_{inf} $;
	Sensing and Perception\;
	Update Q-table and Channel-Location Table\;
	Switch back $ c_{inf}=c_{old} $\;
}
/* Policy 3: modified soft-max */ \\
\If{{$u_d>u_{thr} \& \frac{\rho_d}{u_d}\ll 1$} OR {$ \Delta_{retr.,r}>\Delta_{retr._{thr}} $} OR {$\Delta_{err.,r}>\Delta_{err._{thr}} $}}

{   Calculate $\varepsilon(s)$ using Eq. \ref{eq:vareps} \\
	\If{$ Uniform(0,1) \leq \varepsilon(s) $}
	{ 
		$ \rho (s,a)= min \{\rho_o (s,a), \rho_u (s,a)\} $ \;
		$ \rho_{max}=max (\rho (s,a))  $ \;
		$ \rho_{min}=0.9\times\rho_{max} $ \;
		Calculate Euclidean distance for each action $a$ versus current action iff $ \rho (s,a)>\rho_{min}$ and multiply it with channel diversity factor of action $a$, i.e. calculate factor $\kappa_a$ \; 
		$ a^*= argmax \left\{\kappa \right\} $ \;
	}
	\Else
	{
	    $ Q_{max}=max (Q (s,a))  $ \; $ Q_{min}=0.85*Q_{max}$\;
		Calculate Euclidean distance for each action $a$ versus current action iff $ Q (s,a)>Q_{min}$ and multiply it with channel diversity factor of action $a$, i.e. calculate factor $\kappa_a$ \;
		$ a^*= argmax \left\{\kappa\right\} $ \;
	}
}

/* Policy 4: Decision Making - Control Stage */ \\
\If{{$ Q(s,a^*) = 0 $} OR {$ Q(s,a^*) !=0 \& Q(s,a^*)>1.15* q_{curr} $}}
{
	apply action {$a^*$}
}
\Else
{
  keep current configuration 
  {$a^*=NULL$}
} 
Update cumulative reward $ Q $-value using Eq. \ref{eq:qval}\;
}
return {$a^*$}

\end{small}
\end{algorithm}

\section{Performance Evaluation}
\subsection{Simulator Environment}
To evaluate the proposed framework, we use the IEEE 802.11 compliant discrete-event network simulator ns-3. We consider scenarios where there is the mAP in conjunction with a single EXT and a variable number of client devices. The EXT is modeled as a node that has two radios---one, an adhoc mode interface that is used to establish backhaul communication with the mAP, and the second, an AP mode interface that is used to allow client devices to associate.

All subsequent tests were carried out with all the radios operating on the 2.4 GHz band and a channel width of 20 MHz. Packet size is set to 1000 bytes and transmission power of all radios is 12 dBm. SNR based ideal rate adaptation is used and the MAC protocol is IEEE 802.11.



In every test, we transmit a Constant Bit Rate (CBR) UDP data stream of 5 Mbps from the mAP to each of the client devices, and the ICALO parameters are set as: $\varepsilon_{EXT}(0)=1$, $\varepsilon_{mAP}(0)=1$, $Temp=50$, $\sigma=100$, $\psi_{EXT}=\frac{1}{121}$, $\psi_{mAP}=\frac{1}{11}$, $\eta=0.7$, $\gamma=0$, $\tau=2$, $u_{thr}=60 (\%)$, $RSSI^{\prime}=-60\text{ dBm}$, $\Delta_{err.}=0.005\%$, $\Delta_{retr.}=50\%$.

The network area considered was 20 m $\times$ 10 m. All the nodes of our network (mAP, EXT and client devices) are placed within this area. For the purpose of this simulation, we may consider it as an apartment of length 20 m and width 10 m, consisting of 8 rooms as given in Fig. \ref{fig:convergence_graph} (a). Additionally, APs belonging to neighboring external networks may be placed outside of this network area. In all tests, every node (internal or external) was placed in an enclosing area of 30 m $\times$ 20 m. Note that all interfering external APs transmit at a rate of 5 Mbps to an assosciated node placed outside the apartment.

We divide the testing process into three phases to highlight different aspects of our approach.
1) Speed of convergence to near-optimal throughput
2) Comparison of steady-state throughput to other channel-assignment schemes
3) Resilience to dynamic network conditions.
\subsection{ns-3 Simulation Results}

\subsubsection{Speed of convergence to near-optimal throughput}

ICALO takes time in trying out different channel assignments and locations before arriving at a final state (steady-state). Therefore, it is important to understand how the network throughput will be affected during this period. In this experiment, we try to demonstrate this behavior of ICALO and more importantly, its speed of convergence to the steady-state throughput. Short time periods for channel switching and repositioning are not considered in the analysis of results as they make no effect on the convergence behavior (other than act as small delays).

\begin{figure}
\centering
    \subfigure[Simulation environment for demonstrating convergence]
    {
    \includegraphics[width=0.4\textwidth]{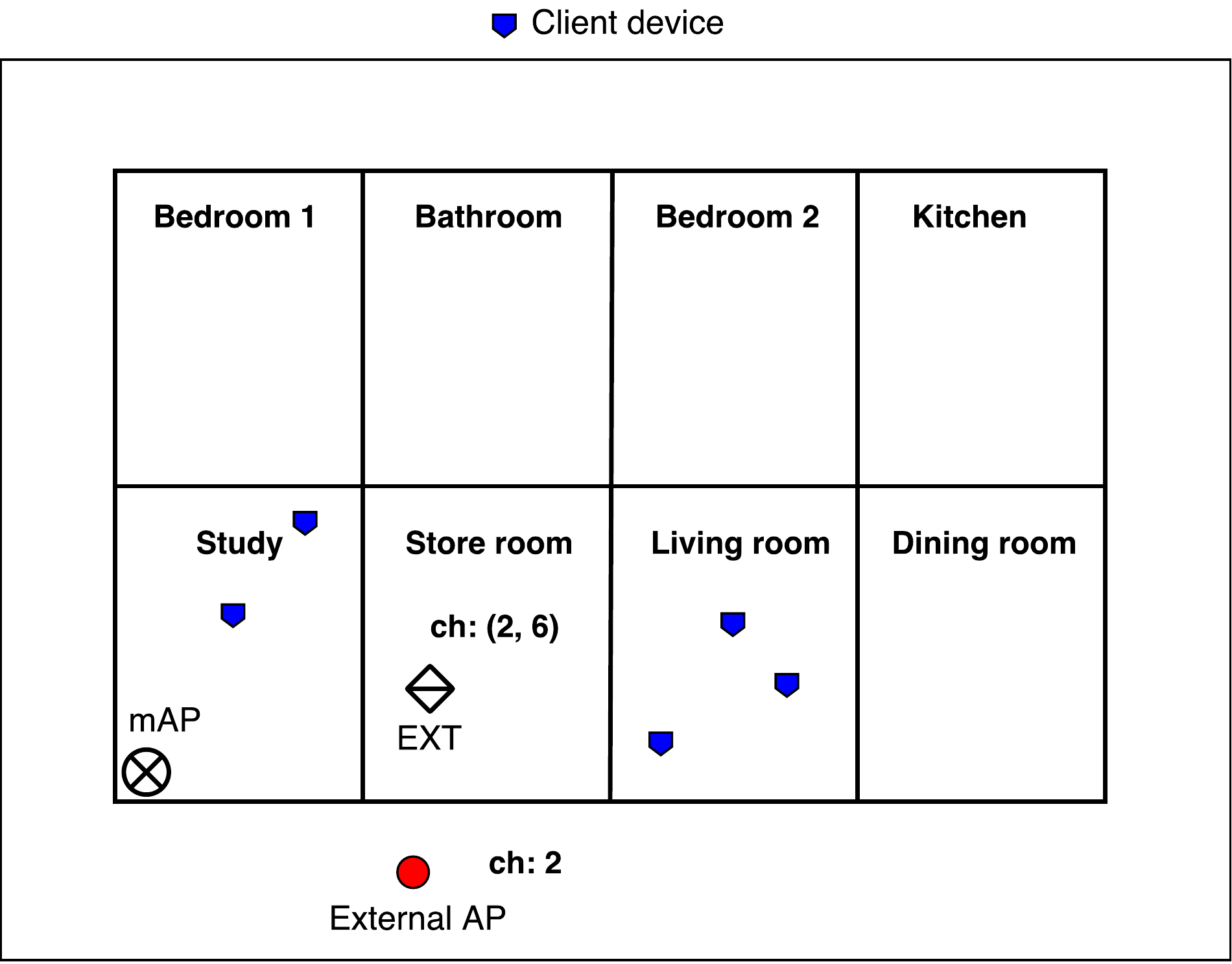}
    }
    \subfigure[Convergence of ICALO to steady-state throughput]
    {
    \includegraphics[width=0.7\linewidth]{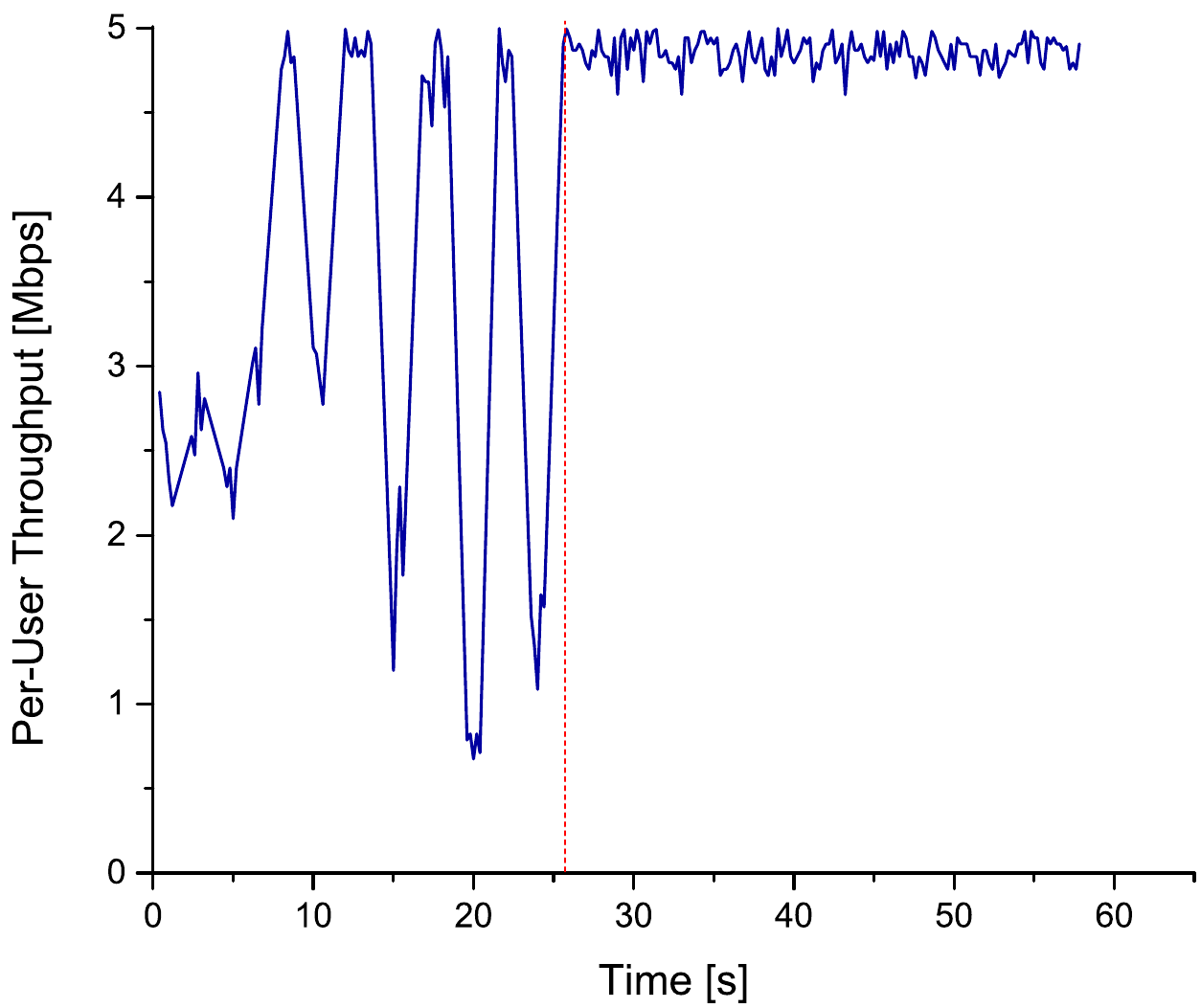}
    }
    
\caption{Comparative analysis for setup in Experiment 1.}
\label{fig:convergence_graph}
\end{figure}

We consider a family of five living in the apartment, three in the living room and two in the study. This is visualized in Fig. \ref{fig:convergence_graph} (a). The mAP and EXT are initially placed as indicated. Then we introduce a single external (non-managed) node to act as external interference to our network. The FH channel and BH channel of the EXT is initially set to 2 and 6, respectively. The external AP channel is set to 2. The variation of per-user throughput versus time when running ICALO in this scenario is given in Fig. \ref{fig:convergence_graph} (b).



%
%
%

Here, we see that the network reaches near-optimal throughput at its steady-state in around 26s, the convergence time of ICALO (indicated by the dashed line). Note that this is just the initial convergence time; as ICALO learns, the convergence time will drop (seen later in Fig. \ref{fig:resilience_graph} (b)). From the more pronounced peaks and valleys before the dashed line in Fig. \ref{fig:convergence_graph} (b), we can get an idea of how many changes in the channel configuration of the EXT occurred before the steady-state (for this particular arrangement of nodes, there was no EXT repositioning suggested by ICALO). The actual number of channel changes to reach the steady-state was 11.

To get a more general idea on the convergence times and the number of configuration changes (location changes plus channel changes) to reach steady-state, we conduct 50 tests with the same configuration as in Fig. \ref{fig:convergence_graph} (a), except that the EXT is placed at a random location within the apartment in each test. The FH and BH channels of the EXT is set to 3 and 7, respectively, while the external AP is in channel 3.


We observed that for 90\% of tests, the convergence time is less than 36 s---all tests converged within 43 s. The mean convergence time was 23.6 s with a standard deviation of 7.9 s. Analysis of configuration changes until steady-state revealed that the mean number of configuration changes is 9.4 and that the standard deviation is 2.5. These numbers further validate what we observed in Fig. \ref{fig:convergence_graph} (b). 

\subsubsection{Comparison of steady-state throughput to state-of-the-art schemes}
In this section, we compare the steady-state throughput of ICALO with that of three other channel assignment approaches---namely single channel assignment, Common Channel Assignment (CCA) \cite{Mar05} and Connected Low Interference Channel Assignment (CLICA) \cite{Adya04} for two different scenarios.
In each scenario, we place the client devices in a constant arrangement of locations, and randomly change the initial position of the EXT 50 times and measure the steady-state per-user throughput of ICALO along with that of the other approaches. Hence, each experiment consists of 200 tests---50 for each approach. Note that within a given experiment, the initial location of the EXT, the channels of the interfering external APs, and the positions of the client devices are kept constant, so as to facilitate a fair comparison.

In single channel assignment, it is to be assumed that there is a single available channel. In each individual test, we assign all possible channels to the EXT fronthaul and backhaul, and consider the throughput of the channel that produced the highest throughput as the throughput for that test. This is to get the best possible throughput for each test under the constraint of using a single channel. The essence of CCA is to assign the same set of channels for each radio of every node in a WMN, to have the maximum possible level of inter-node connectivity while having channel variation to reduce interference. To get a high throughput under this premise while maintaining fairness, we assign a random couple of orthogonal channels to the EXT fronthaul and backhaul in each test. To construct the conflict graph in CLICA, we consider the physical model, which assigns  edge weights based on the value of certain network physical parameters as presented in \cite{Adya04} and originally proposed in \cite{Jain03} (the alternative protocol model does not capture interference due to overlapping channels). 

For the first experiment (Experiment 1), we place the client devices and the external AP in the same positions as that in Fig. \ref{fig:convergence_graph} (a). The channel of the external AP is set to 3 and the initial channel of the EXT fronthaul and backhaul is set to 3 and 7, respectively. Then for each test in the experiment, the position of the EXT is set to a random position within the confines of the apartment. In the second experiment (Experiment 2), there are two people in the dining room, one in the living room, one in the store room and one in the bathroom.

\begin{figure}
\centering
   \subfigure[Simulation environment for Experiment 1]
    {
    \includegraphics[width=0.7\linewidth]{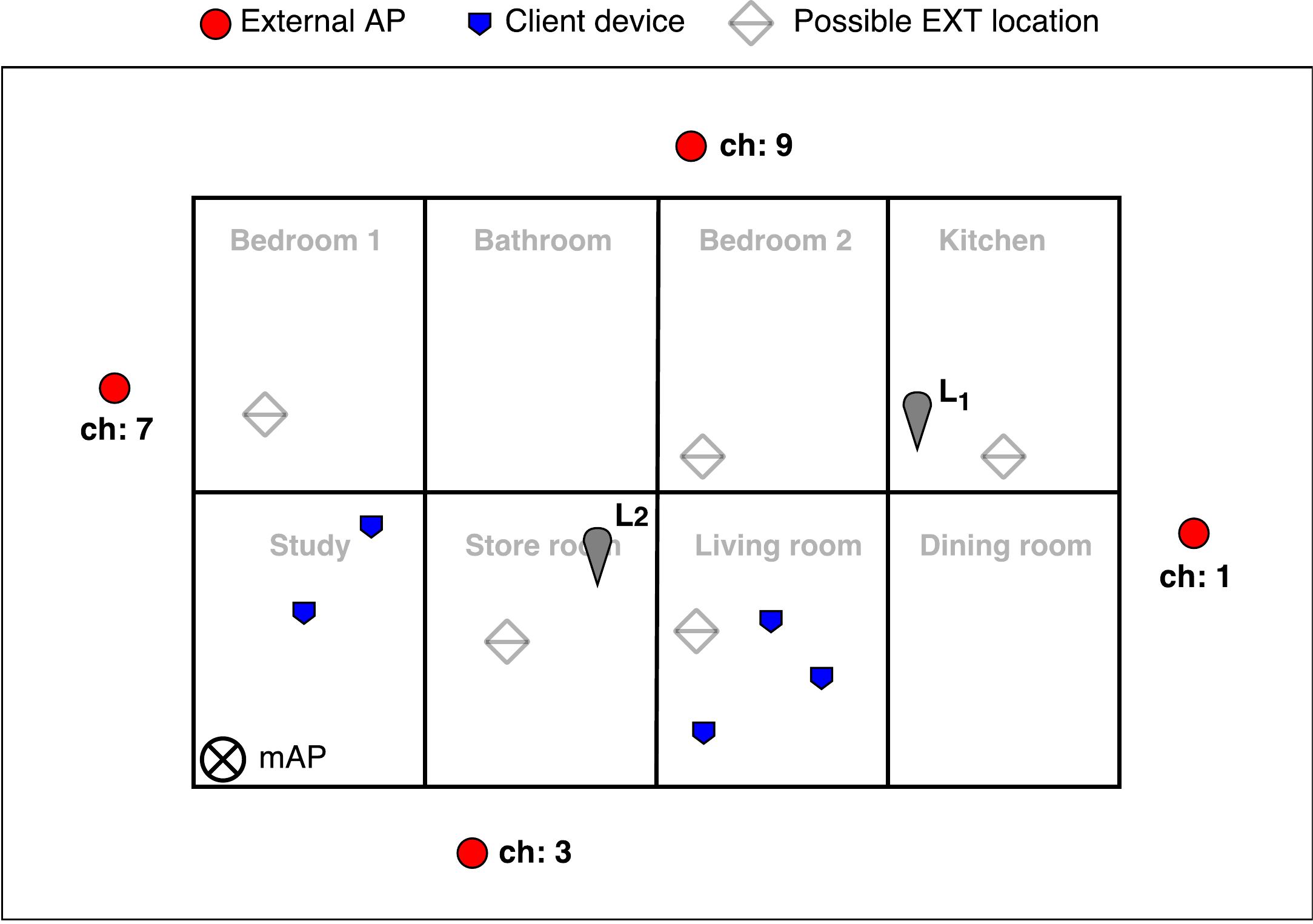}
    }
    \subfigure[Results of Experiment 1]
    {
    \includegraphics[width=0.7\linewidth]{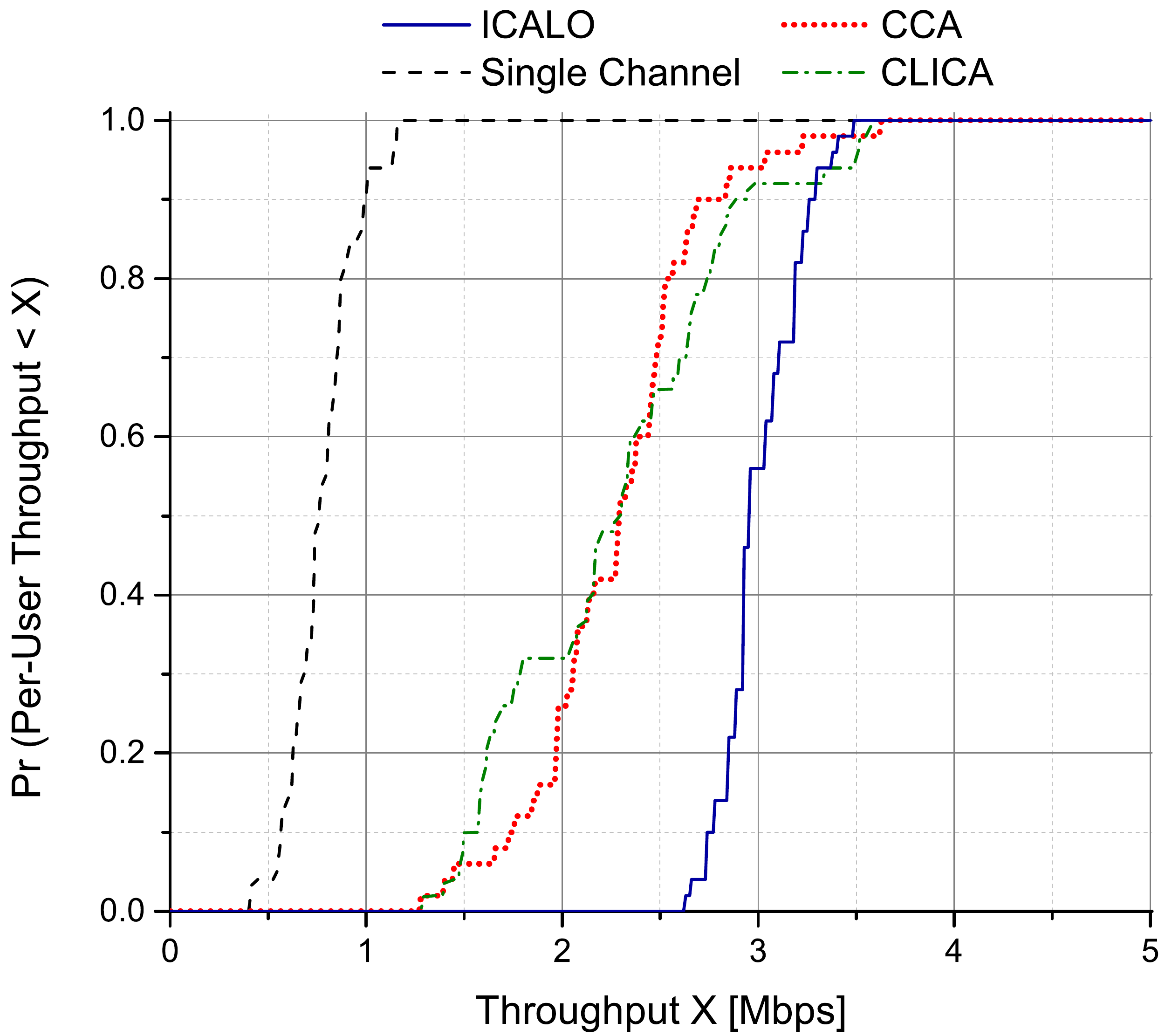}
    }
\caption{Comparing ICALO with single channel, CCA and CLICA for Experiment 2}
\label{fig:compar_exp_2}
\end{figure}

\begin{figure}
\centering
   \subfigure[Simulation environment for Experiment 2]
    {
    \includegraphics[width=0.7\linewidth]{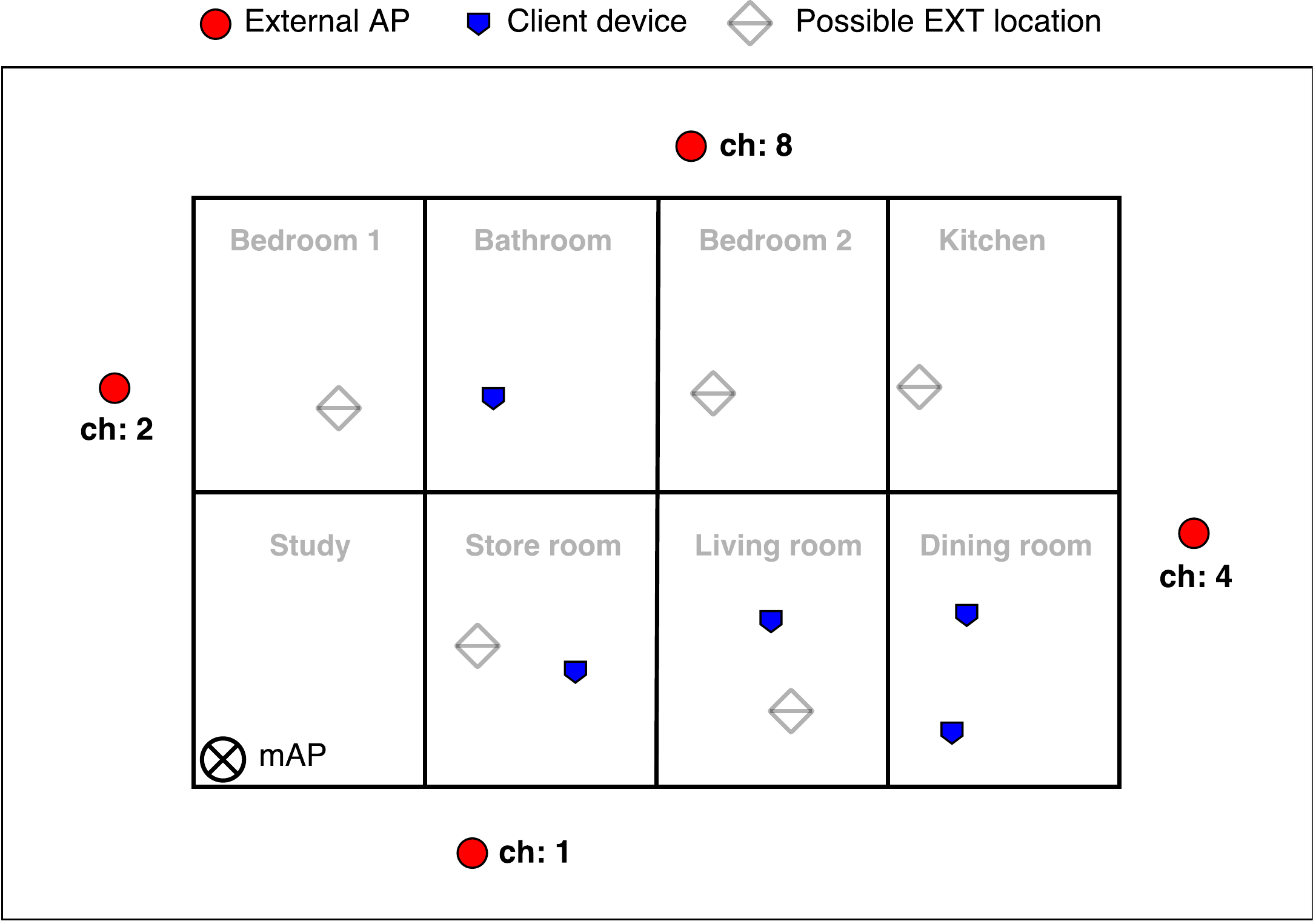}
    }
    \subfigure[Results of Experiment 2]
    {
    \includegraphics[width=0.7\linewidth]{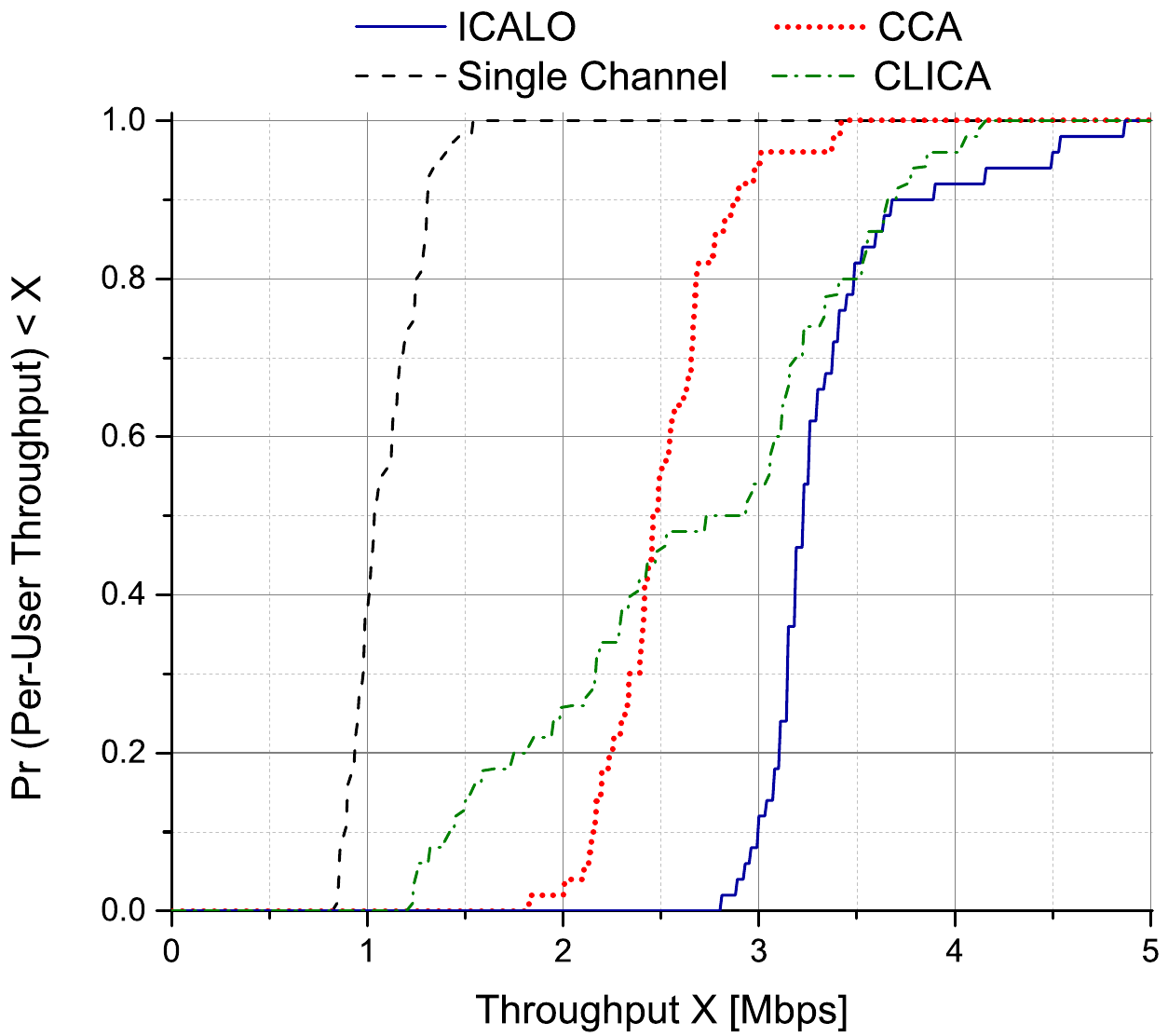}
    }
\caption{Experiment for comparing ICALO performance with single channel, CCA and CLICA - Experiment 2}
\label{fig:compar_exp_3}
\end{figure}

We consider a much more congested environment with four external APs as depicted in Fig. \ref{fig:compar_exp_2} (a) and Fig. \ref{fig:compar_exp_3} (a), respectively. The operating channel of a particular AP is given next to that AP in the figures. In Experiment 1, the respective channels are 3, 7, 9 and 1 while in Experiment 2, they are 1, 2, 8 and 4.


%

%

%
%
%

Results of Experiments 1 and Experiment 2 are portrayed in Fig. \ref{fig:compar_exp_2} (b) and Fig. \ref{fig:compar_exp_3} (b), respectively. For each experiment, the Cumulative Distribution Function (CDF) of the steady-state throughput of each algorithm is plotted. With the increased and more complex external interference, the average steady-state throughput is much less than that of the case of a single interfering channel (\ref{fig:convergence_graph} (a)). As expected, single channel assignment has the worst
overall performance. CCA performs much better as it eliminates (in our tests) inter-channel interference
by choosing orthogonal channels. However, even in this case, it has zero sense of external interference and is inferior to CLICA. CLICA performs better than both the first two approaches, and in some cases, matches the performance of ICALO. But any single formula (as used in CLICA to estimate channel conflicts) is unlikely to fully capture both external and internal interference effects accurately. This is where the exploratory phase of ICALO comes into effect and results in increased
performance.

Some of the low-throughput results of CCA and CLICA in Fig. \ref{fig:compar_exp_2} (b) and Fig. \ref{fig:compar_exp_3} (b) were caused in scenarios where the EXT was placed in locations too far away or too near to the mAP. In such situations, no channel assignment can recover the degradation of throughput caused due to the poor location of the EXT. ICALO was able to alleviate this by repositioning the EXT. For example, the lowest throughput for CCA and CLICA in Experiment 1 is 1.3 Mbps, where in that test, the EXT was initially positioned in location $L_{1}$---inside the kitchen (Fig. \ref{fig:compar_exp_2} (a))---clearly a bad location for it considering the locations of client devices. By initially repositioning the EXT to location $L_{2}$, ICALO was able to eventually obtain a steady-state throughput of 2.7 Mbps. This portrays the tight coupling between channel assignment and location of EXTs in the goal for throughput enhancement. 

\subsubsection{Resilience of ICALO to dynamic network conditions}
With the use of Wi-Fi enabled devices continuing to grow at an astronomical rate, any deployment of a new Wi-Fi network should expect interference from neighboring external networks that is extremely dynamic---new devices will get added and existing devices will leave unpredictably. As such, modern networks should be resilient in the face of these effects and be able to recover from them quickly to reach peak performance. ICALO has a decided advantage in this respect as it keeps getting better as time passes, and is able to make smarter, faster decisions based on its ever-growing knowledge base.

\begin{figure}
\centering
   \subfigure[Simulation environment for experiment]
    {
    \includegraphics[width=0.7\linewidth]{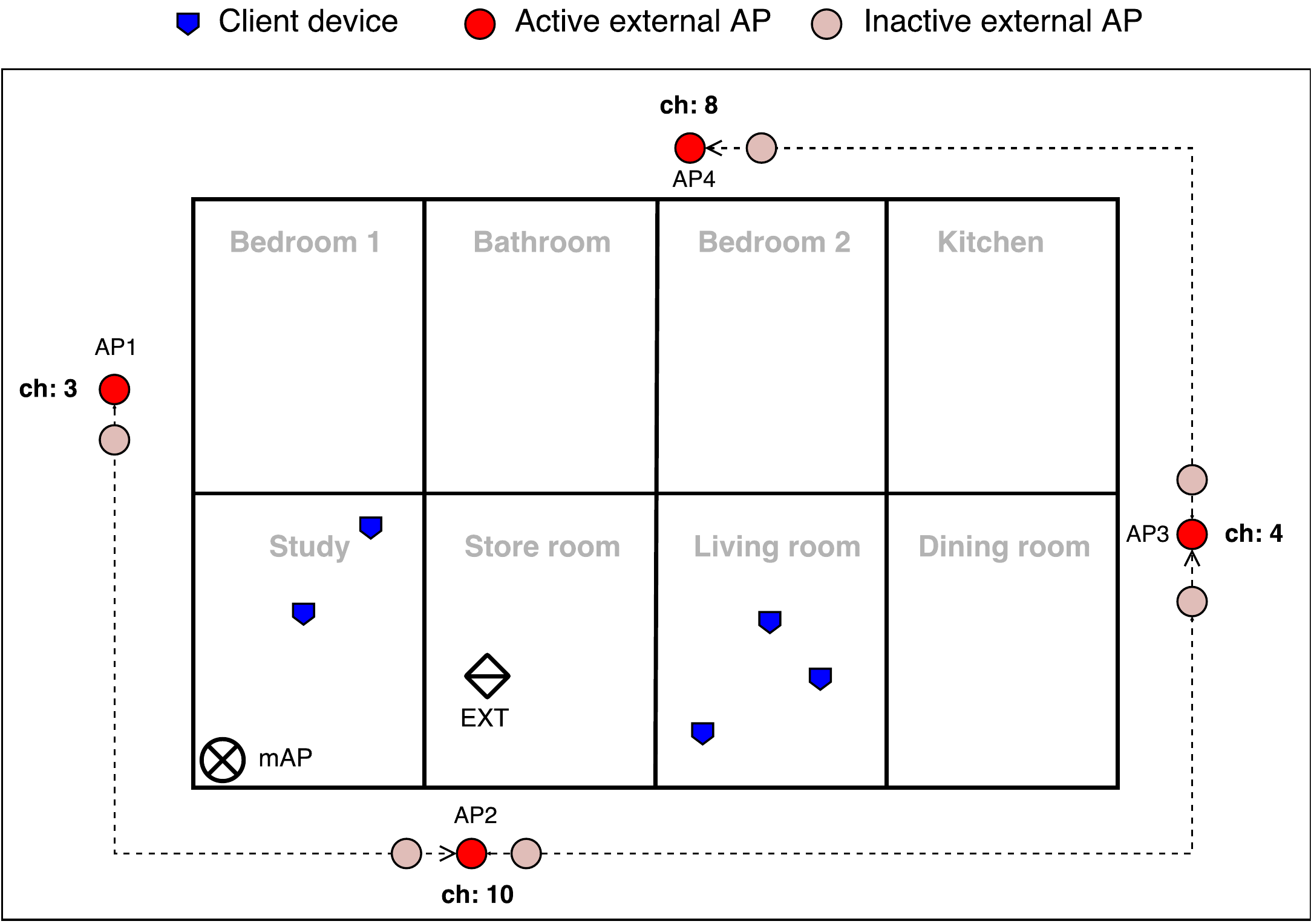}
    }
    \subfigure[Results of experiment]
    {
    \includegraphics[width=0.7\linewidth]{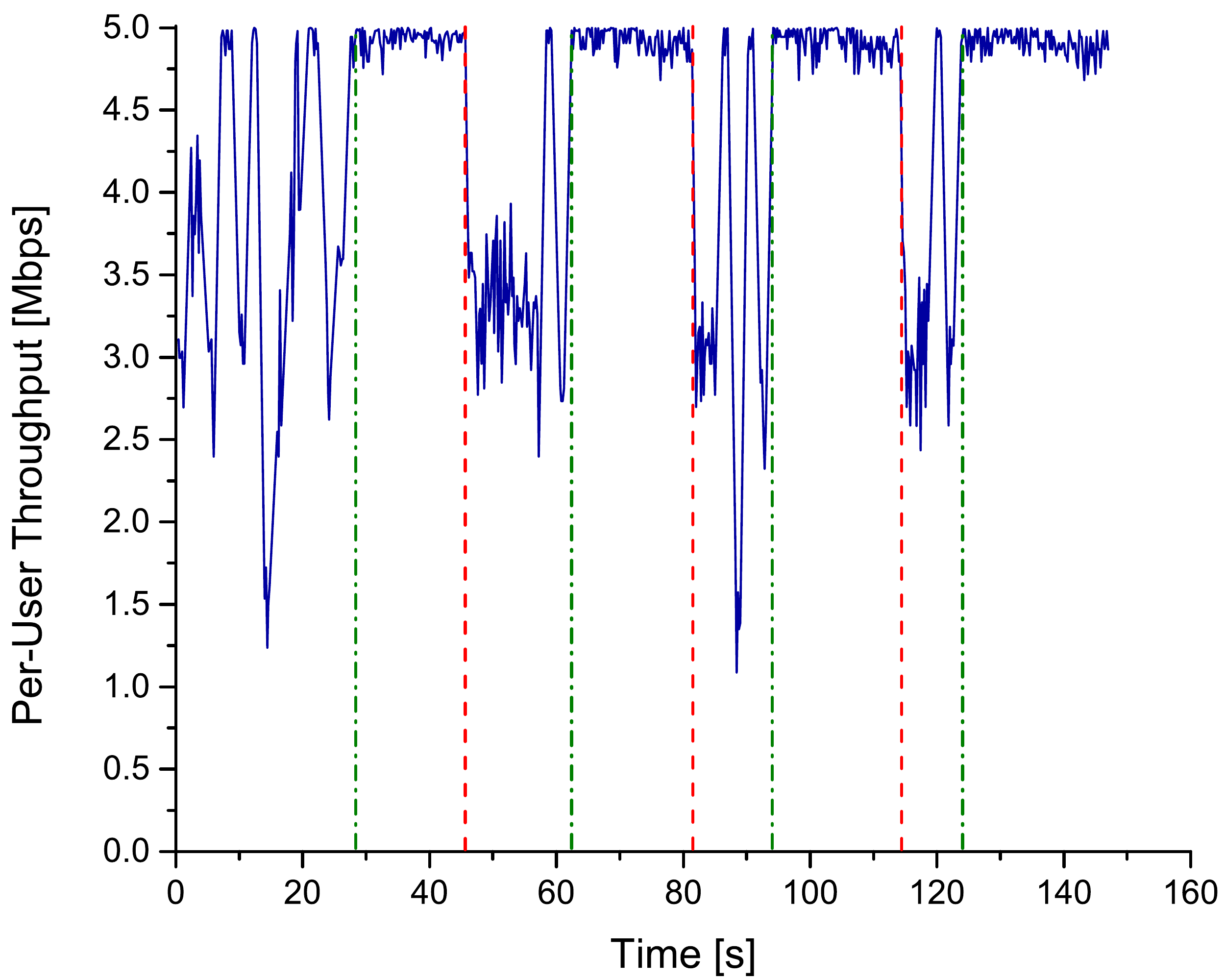}
    }
\caption{Recovery of per-user throughput in dynamic network conditions by ICALO}
\label{fig:resilience_graph}
\end{figure}

To verify this claim, we simulate the scenario illustrated by Fig. \ref{fig:resilience_graph} (a): starting with AP1 activated, remove and activate each of the APs, $\{$AP1, AP2, AP3, AP4$\}$ one by one, pausing for the system to reach a steady-state before removing the current AP and activating the next. Continuing in this order, finally only AP4 is left activated. AP1, AP2, AP3 and AP4 transmit at channels 3, 10, 4 and 8, respectively.

%

%

The per-user throughput variation versus time for this scenario is given in Fig. \ref{fig:resilience_graph}. The moments at which the system reaches the steady-state is marked by dot-dashed lines (green) and the moments at which the current external AP is removed and the next one is activated are marked by dashed lines (red).

It can be seen clearly that the network reaches near-optimal throughput at each stage after successfully recovering from the decline in throughput due to a sudden change in the external interference conditions. Note also that successive convergence times decrease as 28 s, 16 s, 12.5 s and 10 s. This reflects the effect of the growing knowledge base. Following this pattern, as the system evolves, we can ideally expect ICALO to make optimal decisions with little lag.

\section{Testbed Evaluation}
We practically evaluate the feasibility of the proposed framework by developing a full prototype.

\subsection{Testbed Environment}
We consider an experiment with COTS devices running Linux Embedded Development Environment (LEDE) \cite{lede}. For the mAP, we use a TP-Link Archer 1750AC router with one 2.4 GHz radio configured to operate in two modes---ad-hoc (connecting EXT) and infrastructure (connecting user devices). We design one EXT with two 2.4 GHz radios by combining two APs (TP-Link Archer 1750AC) in such a way that LAN interface of one is connected to the WAN interface of other. One AP operates in ad-hoc mode, while the other operates in infrastructure mode. Both the mAP and EXT are having a channel width of 20 MHz. The wireless repeating mode of Wireless Distribution System (WDS) is used to connect mAP and EXT. To test in a more challenging environment, we selected the 2.4 GHz band due to a larger number of neighbors that are not available on the 5 GHz band. We equip the EXT with a USB-to-audio adapter and speaker in order to enable cyber-user interface. By this interface, ICALO notifies an end-user when to re-position the EXT. At both the mAP and EXT, we host a part of the sensing logic which periodically reports network parameters, that is done by combination of Linux Shell and Python programming. The logic of the other blocks of ICALO is hosted on a MATLAB server that uses secure shell (SSH) to push new configurations to the Wi-Fi system.

\subsection{Single mAP with single EXT scenario}
A validation of ICALO is done in the non-managed environment with a layout shown in Fig. \ref{fig:layout}, where mAP is placed at location $l_0$. The initial location of EXT is not pre-defined (ICALO will suggest one). We consider the worst-case scenario with always active users in single and multi-user scenarios. In case of a single-user scenario, the user is located at $l_2$ with 2K video demand. The RSSI from mAP at location $l_2$ is below -75 dBm and to serve this user, an extender is needed. In single-user scenario, ICALO firstly optimizes the location of the EXT and then searches for an optimal channel assignment. With regards to channel assignment, we compare ICALO (with the proposed G-RL agent) with an unguided RL (UG-RL) agent. In both cases, when ICALO or UG-RL agent decides on the optimal channel combination, a hidden node is introduced at the backhaul link to test their responsiveness. In the multi-user scenario, the mAP and EXT each have two connected and active user devices. Parameters related to ICALO and the system are listed as: $\varepsilon_{EXT}(0)=1$, $\varepsilon_{mAP}(0)=1$, $Temp=50$, $\sigma=100$, $\psi_{EXT}=\frac{1}{121}$, $\psi_{mAP}=\frac{1}{11}$, $\eta=0.7$, $\gamma=0$, $\tau=4$, $u_{thr}=60 (\%)$, $RSSI^{\prime}=-65\text{ dBm}$, $\Delta_{err.}=0.005\%$, $\Delta_{retr.}=50\%$.



\subsubsection*{Location Optimization}
The initial recommendation of the EXT's placement is mid-way between the locations of user device at $l_2$ and mAP at $l_0$ in Fig. \ref{fig:layout}, somewhere close to location $l_4$. After the initial placement of the EXT, by means of sensing and perception, ICALO validates the average RSSI level of the EXT received at the mAP's location (-70 dBm) and the RSSI level of the user device at the EXT's location $l_4$ (-44 dBm). Since the RSSI level of the EXT received at mAP is below $RSSI^{\prime}=-65\text{ dBm}$, ICALO sends a voice notification to the user to reposition the EXT to a new location $l_1$, mid-way between the EXT's current location, $l_4$, and the mAP's location, $l_0$. After re-positioning the EXT to $l_1$, ICALO again validates the average RSSI level of the EXT received at the mAP's location $l_0$ (-56 dBm) and RSSI level of the user device at the EXT's current location $l_1$ (-58 dBm). Since the RSSI levels satisfy the RSSI constraint, ICALO can start searching for an optimal channel combination for BH and FH links.

\subsubsection*{Unguided Channel Optimization}
Unguided channel optimization relies only on BSmax probabilities without domain knowledge. That is, when $\xi<\varepsilon (s)$, only $\rho_{o}$ is considered when selecting the next action (this is the classic Softmax exploration). As a consequence, the UG-RL agent requires a longer searching time to find an optimal configuration as illustrated in Fig. \ref{fig:Res1}(a), with a high likelihood to apply channel combinations with poor performance. Thus, the Wi-Fi system experienced poor performance for a longer time in comparison with ICALO. Also, to find optimal channel combinations, UG-RL agent applies far more actions (higher learning cost) than ICALO, leading to the degradation of user experience due to many re-connections and delays for re-association of both EXT's and user devices. We note here that the channel combinations with poor performance due to high level of contention, and/or large errors caused by hidden nodes, require more time to establish connection between mAP and EXT, and also between user devices and mAP/EXT. This time (in range of several to tens of minutes) is referred to as a dead time in the Wi-Fi system, and it increases with higher channel utilization and/or interference. To reduce the dead time, a distributed logic at both the EXT and mAP is added (EXT is not visible to G-RL agent in the cloud) to reset the system configuration to so far best-known settings. As such, ICALO has a much smaller probability to visit actions with poor performance compared with a UG-RL agent.

\subsubsection*{Guided Channel Optimization}
\begin{figure}
\centering
   \subfigure[Instant Reward (UG-RL agent)]
    {
    \includegraphics[width=0.7\linewidth]{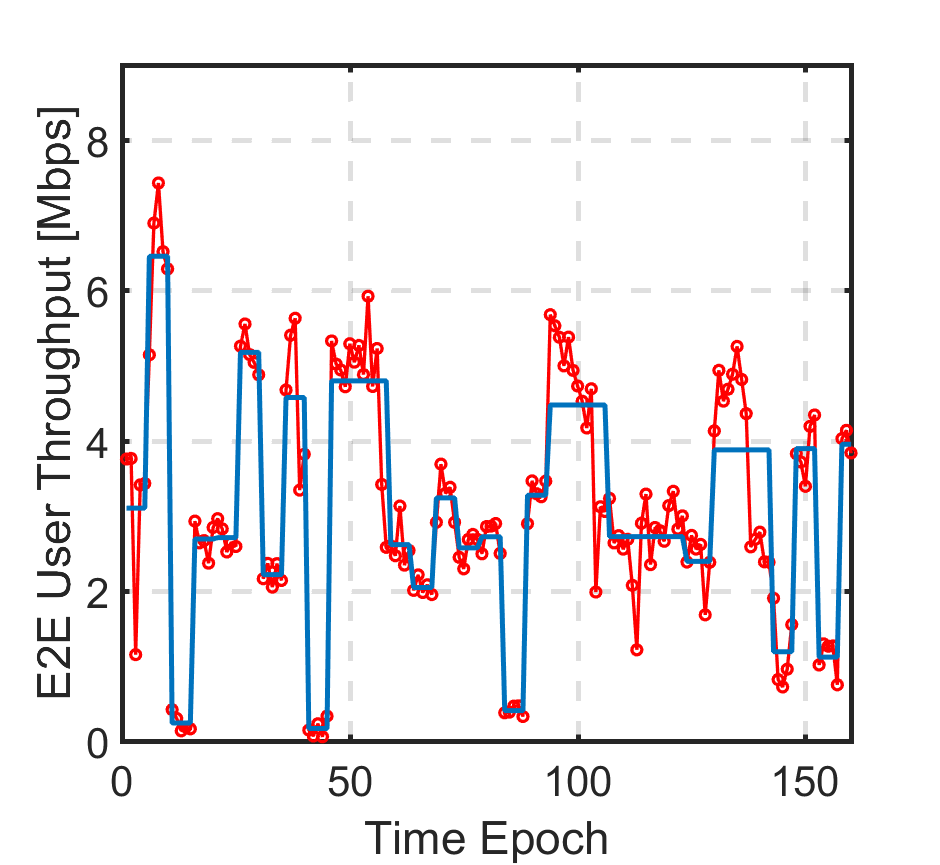}
    }
    \subfigure[Instant Reward (G-RL agent)]
    {
    \includegraphics[width=0.7\linewidth]{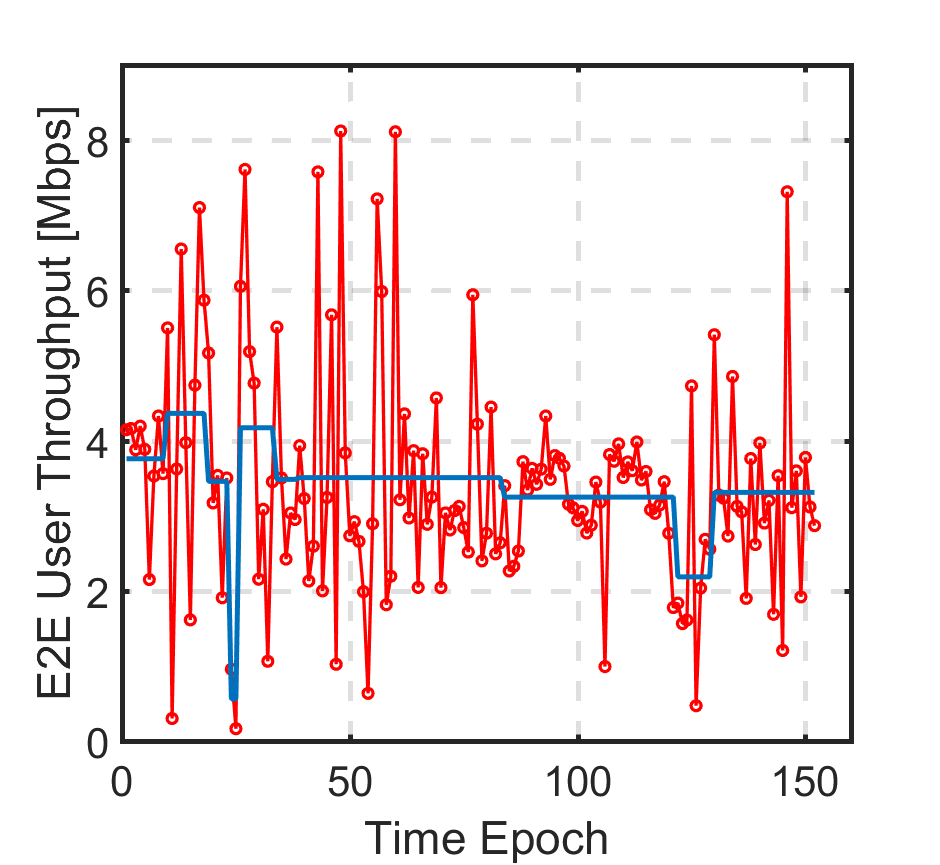}
    }
\caption{Single-User Scenario}
\label{fig:Res1}
\end{figure}

Guided by the domain knowledge, the G-RL agent used in ICALO significantly decreases the search time for the optimal channel configuration as shown in Fig. \ref{fig:Res1} (b). To maximize the initial learning space of ICALO, the agent starts with non-overlapping channels (e.g. channel 3 for BH and channel 8 for FH). After a new configuration is pushed, ICALO collects a number of sensing samples (4 in our case) with period $T=4$ s before reasoning about applying a new action. Also, to avoid situations where a collection of a certain number of sensing samples lasts long, ICALO specifies the maximal time it will wait for collection as 120 s. Subsequently, with each new channel action, ICALO sends to the node the best-known channel action so far. This is necessary to avoid the channels which don't allow re-establishment of all links of a certain node in 30 s. For those channels, ICALO sets the channel utilization to 1000, to stress poor performance at those channels. The tested environment includes 61 and 72 non-managed neighbors sensed at mAP's location and EXT's location, respectively. The level of contention is very high for each channel and most of the channels are highly utilized (see Table \ref{table:ch_loc_table}, which is updated by the perception block with each new sensing sample). By applying a new action (channel configuration), ICALO acquires knowledge about the utilization of the current and adjacent channels, and calculates hidden node and contention node impacts. In that way, ICALO keeps the average throughput in the Wi-Fi system approximately constant (4 Mbps) and only needs a very short period ($T=40$ s) to learn the neighborhood. From $T=40$ s, G-RL agent chooses between two channel combinations (4,10) and (3,7). After the G-RL agents stabilizes the Wi-Fi system, we add a hidden node at the backhaul at time instant 120 s, to test ICALO's responsiveness to a dynamic environment. An additional AP is placed at location $l_3$ which operates at channel 3 and has an associated active user device at saturated traffic load. Consequently, the utilization of channel 3 is increased from 38 to 73. Here, ICALO detected a very high level of error rate at the mAP, and takes only two iterations to avoid the hidden node problem as illustrated in Fig. \ref{fig:Res1}(b). 

\begin{table}
\begin{center}
\begin{tabular} {|c|c|c|c|c|c|c|c|c|c|c|c|}

\hline
Loc/Ch& 1&2&3&4&5&6&7&8&9&10&11 \\ 
\hline
$l_{mAP}$& -&1e3&38&39&	38&	50&	-	&40	&74&	-&	1e3 \\ 
\hline
$l_{EXT}$& 62&1e3&37&37	&61	&84&	71	&35	&74	&44	&1e3 \\ 
\hline

\end{tabular}
\end{center}
\caption{Channel - Location table}
\label{table:ch_loc_table}
\end{table}

\subsubsection*{Multi-users scenario}
\begin{figure}
	\centering	
	{\includegraphics[width=0.7\linewidth,trim={0cm 0cm 0cm 0cm},clip]{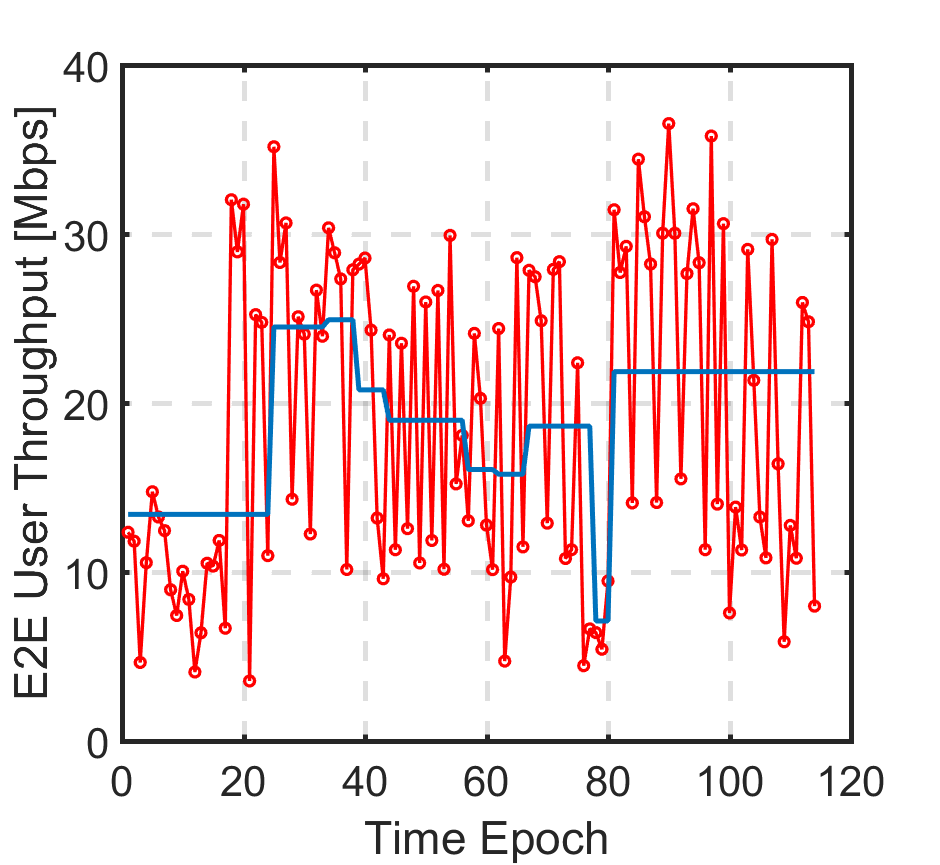}}
\caption{Multi-Users Scenario (G-RL agent)}
	\label{fig:Res4}
\end{figure}

In this scenario, we consider only  ICALO (G-RL). There are 2 user devices connected to the mAP with RSSI levels of -35 dBm and -56 dBm and 2 user devices connected to EXT with RSSI levels -48 dBm and -58 dBm. All the devices stream a 2K video. As shown in Fig. \ref{fig:Res4}, ICALO performs very well in case with multiple users and avoids the channels with poor performance. It is worth noting that the tested environment is very dynamic, and during testing we observed that ICALO very quickly adapts to changes in the neighborhood.

%
%

%
%
%

%

\section{Conclusion and Future Works}\label{sec:Conclusion}
This paper presented ICALO, a self-optimization scheme for wireless extenders in a WMN which adopts an AI-driven learning framework. ICALO optimizes the operating channels and locations of extenders by striking a balanced trade-off between their backhaul and fronthaul performance, considering the impact of uncoordinated neighboring networks, learning cost and network dynamics. Our results show significant throughput improvements over several other channel assignment approaches while converging to peak-performance much faster and with a lower number of actions than a UG-RL. We also portrayed the resilience of ICALO by demonstrating its ability to quickly recover from throughput degradation caused by sudden changes in the network environment. We relate this performance to the guidance of the reinforcement learning agent by using domain knowledge to curb unnecessary exploration while fostering smarter exploitation. We conclude that ICALO successfully addresses the NP-hard problem of joint channel assignment and location optimization of WMNs by guaranteeing low-cost learning and achieving near-optimal network configurations.

The basic idea of the proposed ICALO framework is validated for a 2-hop network with both single and multiple stations by considering only the 2.4 GHz band. In most modern households, a single extender coupled with an mAP will cover the performance needs of an overwhelming majority of users. As such, the demonstrated performance of ICALO will serve as an important marker of the Quality of Service that can immediately be offered to these users through home wireless networks. However, as a future work, we aim to thoroughly evaluate the performance of ICALO in more complex scenarios such as in WMNs with multiple extenders with dual-band radios (2.4 GHz and 5 GHz).
 

\section*{Appendix A}\label{appendix}
Proof of Lemma 1 is given below.
\begin{proof}
Under the assumption that the location of each node $v_i \in V$ is already determined, and given the neighboring environment, then one sample of our problem can be described as $V=\{v_0, v_1, \ldots, v_{M+U} \}$. We assume $E=\{e_{ij}\mid t_c(v_i,v_j)\neq\varnothing\}$, where $ t_c(v_i,v_j)$ denotes the channel constraints matrix for the nodes $v_i$ and $v_j$. For example, the channel constraints matrix contains the connectivity constraints mentioned above. Defined in such a way, $G=(V,E)$ presents an instance of an NP-hard coloring problem \cite{CPNPhard}. An optimal coloring of $G$ given by $C \times V \rightarrow \{1, \cdots, X(G)\}$ is also an optimal channel assignment for the set $V$ under the channel constraints matrix, already given a set of extenders' locations and a static environment. Other set of extenders'  locations and other instances of the environment might result in different $X(G)$. $X(G)$ denotes the minimal number of colors necessary to color the nodes of $G$ such that no two adjacent nodes receive the same color. In the coloring problem, the coloring is equivalent to channel assignment, thus a color means a channel index.  On the other hand, if $G=(V,E)$  is an instance of the coloring problem and we let $V^{\prime}=\{v_0^\prime, v_1^\prime, \ldots, v_{M+U}^\prime\}$ and $t_c(v_i^\prime,v_j^\prime)$, where $t_c(v_i^\prime,v_j^\prime)=\{0\}$ if $\{e_{ij}^\prime\}\in E$ or $t_c(v_i^\prime,v_j^\prime)=\{\}$ if $\{e_{ij}^\prime\} \not\in E$ ($\{0\}$ denotes non-empty set). Now, if an optimal channel assignment for $V^\prime$ is given by $C^\prime \times V^\prime \rightarrow \{1, \cdots, min(V^\prime,t_c)\}$, then $C\prime$ is also an optimal coloring for $G$, i.e. $X(G)=min(V^\prime,t_c)$ \cite{cpequivalents}. Here, $t_c$ is the new channel constraint matrix and $min(V^\prime,t_c)$ is a minimum-order channel assignment for $V^\prime$. 

Since the formulation of our self-optimization problem is equivalent to the colouring problem (with constraints of static environment and given EXT locations), we deduce that the defined problem is NP-hard.
\end{proof}



%

\end{document}